\documentclass[pra,superscriptaddress,aps,showkeys]{revtex4}
\usepackage{amsmath}
\usepackage{hyperref}
\usepackage{subfigure}
\usepackage{graphicx}
\usepackage{amssymb}
\usepackage{amsfonts}
\usepackage[normalem]{ulem}
\usepackage{xcolor}

\newenvironment{proof}[1][Proof]{\textit{#1:} }{ $\square$}

\usepackage{srcltx}
\begin{document}
\providecommand{\keywords}[1]
{
 \small 
 \textbf{\textit{Keywords---}} #1
 }

	\newcommand{\be}{\begin{equation}}
	\newcommand{\ee}{\end{equation}}
	\newtheorem{corollary}{Corollary}[section]
	\newtheorem{remark}{Remark}[section]
	\newtheorem{definition}{Definition}[section]
	\newtheorem{theorem}{Theorem}[section]
	\newtheorem{proposition}{Proposition}[section]
	\newtheorem{lemma}{Lemma}[section]
	\newtheorem{help1}{Example}[section]
\renewcommand{\theequation}{\arabic{section}.\arabic{equation}}

\title{Nonlinear lattices from the physics of ecosystems: The Lefever-Lejeune nonlinear lattice in $\mathbb{Z}^2$}

\author{Nikos I. Karachalios\footnote{Corresponding Author.}, Antonis Krypotos and Paris Kyriazopoulos }
\affiliation{Department of Mathematics, University of the Thessaly, Lamia, 35100  Lamia, Greece}

\begin{abstract}
We argue that the spatial discretization of the strongly nonlinear Lefever-Lejeune partial differential equation defines a nonlinear lattice that is physically relevant in the context of the nonlinear physics of ecosystems, modelling the dynamics of vegetation densities in dry lands.  We study the system in the lattice $\mathbb{Z}^2$, which is especially relevant because of its natural dimension for the emergence of pattern formation. Theoretical results identify parametric regimes for the system that distinguish between extinction and potential convergence to non-trivial states. Importantly, we analytically identify conditions for Turing instability, detecting thresholds on the discretization parameter for the manifestation of this mechanism. Numerical simulations reveal the sharpness of the analytical conditions for instability and illustrate the rich potential for pattern formation even in the strongly discrete regime, emphasizing the importance of the interplay between higher dimensionality and discreteness.
\end{abstract}


\maketitle

\section{Introduction}

In the present paper, we pursue further our studies initiated in \cite{LLPhysD} for the Lefever-Lejuene  nonlinear lattice associated to the  exciting theme of the nonlinear physics of ecosystems \cite{EhudBook}, and particularly, to vegetation pattern formation process, \cite{LL1997}, \cite{LL1999JV}, \cite{LLPRE2002}, \cite{LL2004QC}, \cite{Tlidi2008}, \cite{Bordeu2016}. Herein, we will investigate the more relevant and of exceptional interest case of the {\em higher-dimensional lattice}, as exciting pattern formation dynamics may emerge  due to the interplay of discreteness and higher-dimensionality, see \cite{hiD2,hiD1}. For instance, the question we will investigate, is {\em in what extend a transition sequence in the morphologies of the potential 2D-patterns can be alerted in the discrete regime, as in the continuous limit} \cite{LL1999JV},\cite{LL2004QC}.

In the case of the $2D$-lattice $\mathbb{Z}^2$, the discrete Lefever-Lejeune equation (DLL) is given by the system of equations
\begin{eqnarray}
\label{eq1}
\dot{U}_{n,m}+\frac{\gamma_1}{h^4}U_{n,m}\Delta_d^2U_{n,m}+ \frac{\gamma_2}{h^2}U_{n,m}\Delta_dU_{n,m}-\frac{\gamma_3}{h^2}\Delta_dU_{n,m} -f(U_{n,m})=0.
\end{eqnarray}
In Eq.~(\ref{eq1}),  $U_{n,m}(t)$ is the unknown function occupying the lattice site $(n,m)\in\mathbb{Z}^2$,  $\Delta_d$ stands for the $2D$-discrete Laplacian and $\Delta_d^2$ for the $2D$-discrete biharmonic operator; their definition, in a convenient way for our purposes, and their properties, will be discussed below. The parameters $\gamma_i>0$, $i=1,2,3$, and
$h>0$ is the lattice spacing. The nonlinearity in \eqref{eq1} is of the form
\begin{eqnarray}
	\label{eq2}
	f(U)=\alpha U+\beta U^2-U^3,\;\;\alpha,\beta\in\mathbb{R}.
\end{eqnarray} 

The system \eqref{eq1} will be 
endowed with the initial condition 
\begin{eqnarray}
\label{inc}
U_{n,m}(0)=U^0_{n,m},
\end{eqnarray}
and will be supplemented with suitable boundary conditions, that will be discussed below. 

In the above discrete set-up, the lattice \eqref{eq1} can be viewed as a  discretization of the Lefever-Lejeune (LL) partial differential equation (pde),  which in a non-dimensional form reads as
\begin{equation}
	\label{eq5}
	U_t+\gamma_1U\Delta^2U+\gamma_2U\Delta U-\gamma_3\Delta U-f(U)=0,
\end{equation}
where $\Delta$ and $\Delta^2$ are the Laplacian and the biharmonic operator respectively. Equation \eqref{eq5} is a spatially continuous propagation-inhibition model describing the growth of vegetation density $U(x,y,t)$ in arid or semi-arid areas. It is the formal continuum limit of the lattice \eqref{eq1} as $h\rightarrow 0$. The model \eqref{eq5} relies on a short-range cooperative and long-range competitive spatial mechanism to predict  patterns of vegetation that may emerge in drought environments. The LL model \eqref{eq5}, is derived from a spatially non-local integral-differential equation \cite{LL1997} 
	which involves a continuous redistribution-kernel convoluted with
	a density dependent nonlinearity, that captures the dispersal and spatial interactions of individual plants.
Therefore, the LL pde is a biharmonic approximation of the above kernel-based models \cite{LL1999JV}, \cite{LLPRE2002}, \cite{LL2004QC},  \cite{Tlidi2008}, \cite{Bordeu2016}, 
that adequately replicates the qualitative behavior of plant community systems in a strictly isotropic and homogeneous environment. Numerical and analytical studies in two spatial dimensions, have revealed the pattern forming potential of the spatially continuous LL equation \cite{LL1999JV}, \cite{LLPRE2002}, \cite{LL2004QC}, \cite{Tlidi2008}, ranging from localized structures as spots and isolated "circles", to stripes, ``polygonal lattices'' and Turing periodic patterns.  

In the above physical context, spatially discrete systems can be particularly relevant. They can be used to model vegetation patterns dynamics in a situation where the spatial domain is subdivided into a discrete number of cells \cite{Borgogno2009} (see also \cite{LLPhysD} and references therein). We argue that  the DLL lattice can describe interacting plant dynamics in the presence of the discrete operators involved. For this purpose, we discuss the particular terms of the  DLL model \eqref{eq1}. The parameter $\alpha =1-\mu$, where $\mu$ represents the mortality to growth rate ratio, 
measures the environment's aridity which characterizes the productivity of the system. The parameter  $\beta=\Lambda-1$, where $\Lambda$ represents the cooperation effect influencing the local reproduction;
it is considered as weak for  $\beta \leq 0$ ($\Lambda\leq 1$) and strong for $\beta>0$ ($\Lambda>1$). 
The nonlinear discrete biharmonic term of strength $\gamma_1$ and the nonlinear discrete Laplacian  term of strength $\gamma_2$ express the long-range competition for resources. The short-range cooperative interplay among plants is expressed by the discrete Laplacian term of strength $\gamma_3$. As in \cite{LL1997}, we set $\gamma_3=\frac{1}{2}l^2$  where $l$ represents the ratio of the facilitative length  $L_F$ to the inhibition length  $L_I$. These are the spatial distances beyond which cooperation and competition effects respectively, become negligible. It is physically meaningful to assume that $L_I>L_F$  ($l<1$), especially for dry-lands.  Also,  according to \cite{LL1997}, in the continuous limit \eqref{eq5}, the normalized space variable $x$ is defined as $X/L_I$.
Hence, in the dimensionless discrete DLL \eqref{eq1}, since $h$ measures the distance between the lattice nodes, if $h_X$ is the distance in real spatial dimensions, we have that  $h=h_X/L_I$.

It is evident that the parameter $h$ is crucial as it can measure the overlap of the zones of influence between individual plant dynamics, that is, the zones of their inhibitory and facilitative interactions, \cite{StollWeiner1}, \cite{StollWeiner2}: competition among neighboring individuals becomes significant when $L_I>h_X/2$ and is limited when $L_I<h_X$. Thus, in  terms of $h$,  a short range competition effect is present when $1<h<2$. On the other hand, since $L_I>L_F$ the facilitative interactions are limited when $h>1$ and when $h<1$ both mechanisms are active and significant as long as $l=L_F/L_I>h/2$.

Therefore, the interplay between discreteness and nonlinearity incorporating the above dependencies on $h$, raises interesting novel questions concerning pattern formation:  for example, {\em can we expect the emergence and the survival of interesting patterns in the discrete or even in the strongly discrete regime?}

We explore such questions both analytically and numerically and the presentation of the results of the paper is as follows: In Section 2, we describe the set-up of the problem when supplementing the  DLL system \eqref{eq1} with a variety of boundary conditions, vanishing at infinity, Dirichlet and periodic. In Section 3, we prove estimates for the solutions and we identify parametric regimes for the lattice which distinguish between extinction and possible nontrivial asymptotic behavior. The first scenario which is described by the convergence of the dynamics to the trivial steady state $U_{n,m}=0$, is of physical importance since it is associated with {\em desertification}. A brief numerical study for the extinction regimes, highlight the relevance and accuracy of the theoretical predictions.  The second scenario may describe  convergence to nontrivial states of the system. In Section 4, we investigate the latter scenario further. More precisely, we investigate the potential emergence of Turing instability \cite{EhudBook} in the DLL system. This form of instability is characterized by the instability of its spatially uniform steady states and, consequently, by pattern formation expressed through the convergence of dynamics to spatially nonuniform equilibria. The key result that significantly distinguishes the analysis of the DLL system from that of its continuous counterpart {\em is the proof of the existence of a threshold value on $h$ for this instability to occur}. The analysis also yields a detailed description of the instability sets (bands), encompassing the unstable wavenumbers of perturbations around the spatially uniform states. In Section 5 we investigate numerically the dynamics in the instability regime. The numerical results illustrate the sharpness of the analytical predictions and the high potential for pattern formation, depending on the aridity parameters, even in the case of a strongly discrete system where facilitation and competition mechanisms are weak. Section 6 summarizes our results and provides a brief plan for further studies, extending the present ideas to other relevant nonlinear lattice models.
\section{Functional set-up: Boundary conditions and phase spaces}
Equation (\ref{eq1}), will be supplemented  with either vanishing conditions in the case of the infinite lattice, and periodic or Dirichlet boundary conditions, which give rise to a finite dimensional system. The former case involves the standard infinite dimensional sequence spaces, while the latter cases are associated with their relevant finite dimensional subspaces.  For  each case of boundary conditions, we will recall the definition and properties of the relevant phase spaces  and the properties of the discrete Laplacian and biharmonic operator in their functional set-up.
\paragraph{Vanishing boundary conditions in an infinite lattice.} In the case where the system \eqref{eq1} is considered in the infinite lattice  $\mathbb{Z}^2$ supplemented with vanishing boundary conditions 
\begin{eqnarray}
\label{vanv}
\lim_{|n|,|m|\rightarrow\infty}U_{n,m}=0, 
\end{eqnarray}
the problem will be considered in the standard infinite dimensional sequence spaces, $\ell^p$, $1\leq p\leq\infty$,
\begin{eqnarray}
	\label{eq6}
	{\ell}^p:=\left\{\left\{U=(U_{n,m})_{n,m}\in\mathbb{Z}\right\}\in\mathbb{R}:\quad
	\|U\|_{\ell^p}:=\left(\sum_{n,m\in\mathbb{Z}}|U_{n,m}|^p\right)^{\frac{1}{p}}<\infty\right\},
\end{eqnarray}
 with the key inclusion properties
\begin{eqnarray}
	\label{eq7imb}
	\ell^q\subset\ell^p,\quad \|U\|_{\ell^p}\leq \|U\|_{\ell^q}, \quad 1\leq q\leq p\leq\infty.
\end{eqnarray}	

We proceed by recalling some useful properties of the operator $\Delta_d$ in higher dimensional lattices $N\geq 1$.  The operator $(\Delta_d U)_{\mathbf{n}}$, $\mathbf{n}\in\mathbb{Z}^N$ is the $N$-dimensional discrete Laplacian
\begin{eqnarray*}
	(\Delta_d U)_{\mathbf{n}}=\sum_{j} (U_{\mathbf{n}+\mathbf{j}}-2 U_{\mathbf{n}}+U_{\mathbf{n}-\mathbf{j}}),
\end{eqnarray*}
where $\mathbf{j}$ are the $N$ unit vectors belonging to the $N$ axes of $\mathbb{Z}^N$. Defined on the Hilbert space $\ell^2$,  $-\Delta_{d}: {\ell}^2 \to {\ell}^2$ is self-adjoint and non-negative, that is, for every $U, W\in\ell^2$,
\begin{eqnarray}
	\langle-\Delta_{d}U,W\rangle_{\ell^2}&=&\langle U,-\Delta_{d}W\rangle_{\ell^2}, \\
0\leq	\langle-\Delta_{d}U,U\rangle_{\ell^2}&\leq& 4 N \sum_{\mathbf{n}\in\mathbb{Z}^N}|U_\mathbf{n}|^2.
\end{eqnarray}
When defined on $\ell^p$, $1\leq p\leq\infty$, the operator $\Delta_d:\ell^p\rightarrow \ell^p$  is continuous, that is there exists a constant $C>0$, such that
\begin{equation}
	\label{genlp8}
	||\Delta_d U||_{\ell^p}\leq C||U||_{\ell^p},\;\;\mbox{for all}\;\; U\in\ell^p.
\end{equation}
The {\em discrete biharmonic operator}
$\left(\Delta_d^2U\right)_{\mathbf{n}}=\left(\Delta_d[\Delta_dU]\right)_{\mathbf{n}}:\ell^p\rightarrow\ell^p$, is also continuous for any $1\leq p\leq\infty$, satisfying 
\begin{eqnarray}
	\label{genlp9}
	||\Delta_d^2 U||_{\ell^p}\leq \hat{C}||U||_{\ell^p},\;\;\mbox{for all}\;\; U\in\ell^p.
\end{eqnarray}
\subsubsection{Finite lattices: Dirichlet and periodic boundary conditions .} 
\begin{figure}[tbh!]
	\includegraphics[scale=0.55]{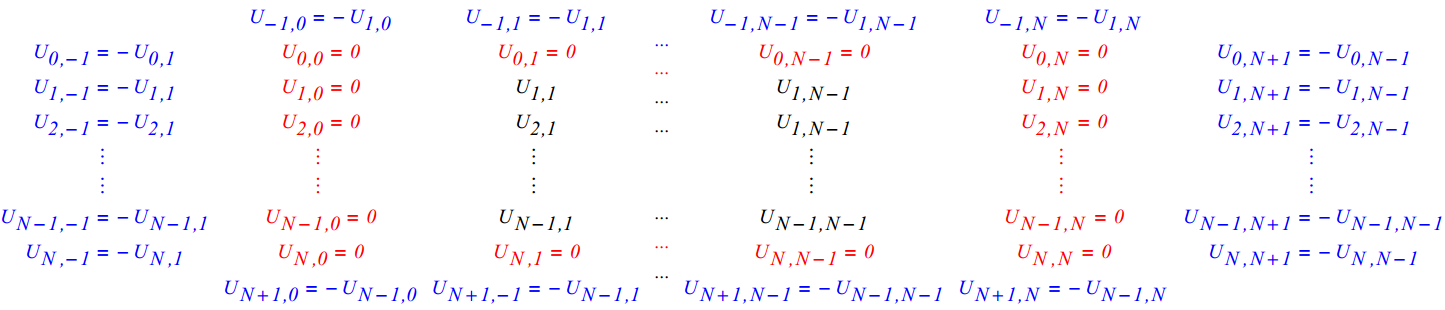}	
	\caption{Implementation of the Dirichlet boundary conditions \ref{DBC} leads to anti-symmetric conditions for the virtual nearest neighbors of the set of the boundary points $\partial\Omega$. Details in the text.}
	\label{fig:D1}
\end{figure}

The finite dimensional dynamical systems of the form \eqref{eq1} arise when the system \eqref{eq1} is endowed with Dirichlet or periodic boundary conditions. We will briefly refer to both cases.
\paragraph{Dirichlet boundary conditions.} To define a finite dimensional lattice dynamical system from Eq.~\eqref{eq1}, we assume that an arbitrary number of $(N+1)^2$ nodes are occupying equidistantly the square interval side $\Omega=[-L,L]\times[-L,L]$, with two-dimensional lattice spacing $h=2L/N$. Accordingly, the discrete spatial coordinates are $(x_n,y_n)=(-L+nh,-L+mh)$, for $n,m= 0,1,2,\ldots,N$, and in Eq.~\eqref{eq1}, we define the function $U_{n,m}(t)=U(x_n,y_m,t)$.  
Regarding the Dirichlet boundary conditions The important feature which should be remarked is that due to the presence of the discrete biharmonic operator the boundary conditions are 
\begin{eqnarray}
\label{DBC}
U_{\mathbf{n}}=0,\;\;(\Delta_d U)_{\mathbf{n}}=0,\;\; \mbox{for all $\mathbf{n}\in \partial \Omega$},\;\;\mathbf{n}=(n,m),\;\;n,m=0,1,2,...,N,
\end{eqnarray}
where $\partial\Omega$ denotes the boundary points of the discretization  of $\Omega$ described above. Due to \eqref{DBC}, {\em virtual} nodes for the nearest neighbors of $\partial\Omega$, $U_{-1,m}$, $U_{N+1,m}$, $U_{n,-1}$ and $U_{n,N+1}$, $n,m=0,...N$ should be introduced in order to make $\Delta_d$ and $\Delta_d^2$ well defined. This way, the implementation of the boundary conditions \eqref{DBC} leads to the anti-symmetric conditions on the virtual nodes as visualized in Figure \ref{fig:D1}; the process is an extension of the one described in detail in \cite{LLPhysD}. 
The simplest case which is natural, as well as, more tractable for the numerical simulations is to presume zero  values for the virtual nodes, that is
\begin{equation}
	\label{DBC2}
	U_{-1,m}=U_{N+1,m}=U_{n,-1}=U_{n,N+1}=0 \;\;\mbox{for all }\;\; n,m= 0,...,N.
\end{equation}
With the boundary conditions \eqref{DBC}-\eqref{DBC2}, the system \eqref{eq1} is considered in the finite dimensional subspaces of $\ell^p$
\begin{eqnarray*}
	\ell^p_0=\Bigg\{U\in\ell^p\;:\;U_{0,m}=U_{N,m}=U_{n,0}=U_{n,N}=0,\;\;n,m,=1,...N\Bigg\}.	
\end{eqnarray*}
The Dirichlet discrete Laplacian as an operator  $-\Delta_d:\ell^2_{0}\rightarrow\ell^2$, has the following eigenvectors and eigenvalues, respectively,
\[
f_{i,j}(n,m)=\sin\left(\frac{ih\pi n}{2L}\right)\sin\left(\frac{jh\pi m}{2L}\right)\mbox{ and }\]
\[ \mu_{i,j}=\frac{4}{h^2}\sin^2\left(\frac{h\pi i}{4L}\right)+\frac{4}{h^2}\sin^2\left(\frac{h\pi j}{4L}\right),\;\;i,j=1,..,N.
\]
Moreover, the following useful, for the derivation of various estimates, discrete version of the Poincar\'{e} inequality holds,
\begin{equation}
\label{crucequiv}
\mu_1\sum_{n,m=0}^{K+1}|U_{n,m}|^2\leq
\frac{1}{h^2}\left<-\Delta_dU,U\right>_{\ell^2}\leq \frac{4}{h^2} \sum_{n,m=0}^{K+1}|U_{n,m}|^2,\;\;\mbox{for every $U\in\ell^p_0$},
\end{equation}
where
\begin{equation}
	\label{prince}
\mu_1:=\mu_{1,1}=\frac{8}{h^2}\sin^2\left(\frac{\pi h}{2l}\right)=\frac{8}{h^2}\sin^2\left(\frac{\pi h}{4L}\right),
\end{equation}
and   $l=2L$ is the length of the symmetric interval  $[-L,L]$.
\paragraph{Periodic boundary conditions.}  In the case of the periodic boundary conditions it can be shown as in \cite{LLPhysD}, that the phase space for the relevant finite dimensional system is the space of periodic sequences
\begin{eqnarray*}
	{\ell}^p_{\mathrm{per}}:=\left\{U=(U_{n,m})_{n,m\in\mathbb{Z}}\in\mathbb{R}:\, U_{n,m}=U_{n+N,m},\, U_{n,m}=U_{n,m+N}, \,
	\|U\|_{\ell^p_{\mathrm{per}}}:=\left(h\sum_{n,m=0}^{N-1}|U_{n,m}|^p\right)^{\frac{1}{p}}<\infty\right\}.
\end{eqnarray*}
We remark that inequalities \eqref{genlp8} and \eqref{genlp9} are valid in the case of the finite dimensional subspaces $\ell^p_0$ of $\ell^2$, as well as in the case of $\ell^2_{\mathrm{per}}$ (see Appendix \ref{AppB} for brief proofs).
\section{Decay of solutions and uniform bounds}
\paragraph{Preliminaries} Decay of solutions, that is $\lim_{t\rightarrow\infty}||U(t)||_{\ell^p}=0$, $1<p\leq\infty$,  is a crucial potential dynamical property for the system \eqref{eq1} since its physical significance is related with the extinction of the vegetation densities and desertification. On the other hand, uniform bounds describing that the solution may not vanish and satisfy  $||U(t)||_{\ell^p}<M$, for some constant $M$, establish  the survival of the vegetation densities and provide as a potential dynamical scenario, the convergence to equilibrium. 

In this section, we discuss both of the above scenarios which depend on certain parametric regimes which will be identified explicitly. Convergence to equilibrium will be discussed in the next section. 

We start by recalling that local existence of solutions can be proved by application of the generalized  Picard-Lindel\"{o}f \cite[Theorem 3.A, pg. 78]{zei85a}. Details are omitted, since the extension of the steps of \cite[Thoerem 2.1 \& Theorem 2.2]{LLPhysD} to the higher dimensional lattices $\mathbb{Z}^N$, $N\geq 1$ it can be easily seen that are  independent of the dimension of the lattice. In the statement of the well-posedness result, we denote by  $Z$  the spaces $\ell^p$, $\ell^p_0$ or $\ell^p_{\mathrm{per}}$, except of the case where it is necessary to refer to one of them, explicitly.
\begin{theorem} 
	\label{thloc}	\begin{enumerate}
		\item (Local existence)
	Assume that $\gamma_i>0$, $i=1,2,3$ and $\alpha,\beta\in\mathbb{R}$, and let $U^0\in Z$, arbitrary. There exists some $T^*(U^0)>0$ such that the initial value problem \eqref{eq1}-\eqref{inc}, has a unique solution $U\in C^1([0,T],Z)$  for all $0<T<T^*(U^0)$. In addition, the following alternatives hold: Either $T^*(U^0)=\infty$ (global existence) or $T^*(U^0)<\infty$ and $\lim_{t\uparrow T^*(U^0)}||U(t)||_{\ell^2}=\infty$ (collapse). Furthermore the solution $U$ depends continuously on the initial condition $U^0\in Z$, with respect to the norm of $C([0,T],Z)$. For all $U^0\in\ell^p$ and $t\in [0,T^*(U^0))$, we may define the map
\begin{eqnarray}
	\label{wds1}
	&&\phi_t:\ell^p\rightarrow\ell^p,\;\;1\leq p\leq\infty,\nonumber\\
	&&\;\;\;\;\;\;U^0\rightarrow \phi_t(U^0)=U(t),
\end{eqnarray}
and $\phi_t(U^0)\in  C^1([0,T^*(U^0)),\ell^p)$. 
\item (Strong continuity of the semiflow)
	Let as assume a sequence $U^{m,0}$, $m\in\mathbb{N}$, converging weakly to the initial condition $U^0$ in $\ell^1$, that is
	\begin{eqnarray}
		\label{weak1}
		U^{m,0}\rightharpoonup U^0\;\;\mbox{in $\ell^1$},\;\;\mbox{as $m\rightarrow\infty$}. 
	\end{eqnarray}
	Then if $T<T^*(U^0)$, we have the strong convergence $\phi_t(U^{m,0})\rightarrow \phi_t(U^0)$ in 	 $C([0,T],\ell^1)$.
	\end{enumerate}
\end{theorem}
It is interesting and important to remark that the strong continuity of the semiflow $\phi_t$ is a consequence of the the {\em Schur property}: weak and norm sequential convergence in $\ell^1$ coincide \cite[Definition 2.3.4 \& Theorem 2.3.6, pg. 32]{Kalton}.

Another novel result we will prove herein, is that solutions starting from non-negative initial data remain non-negative for all times.  The result is of physical significance since (as described in the introductory section) $U_{n,m}(t)$  describes the growth rate of the vegetation density at the point $(n,m)$ of the $\mathbb{Z}^2$-lattice. 
\begin{proposition} \label{posit}
Consider the system \eqref{eq1} supplemented with either case of boundary conditions (vanishing, periodic  
or Dirichlet) and assume that $U_{n,m}(0)\geq0$ for all n,m$\in{N}$. Then $U_{n,m}(t)\geq0$ for all $(n,m)\in\mathbb{Z}^2$ and $t>0$. 
\end{proposition}
\begin{proof}
We argue by contradiction. That is, under the assumption  $U_{n,m}(0)\geq0$ on the initial data, we may assume that due to the smoothness of the solution  $U\in C^1([0,T],Z)$,  there exists a time $t_0>0$ and at least a point $(n_0,m_0)$ of the lattice such that  $U_{n_0,m_0}(t)<0$ for all $t_0<t\leq \tilde{T}_0$, where $\tilde{T}_0<T$, and $U_{n_0,m_0}(t_0)=0$, while the neighboring nodes to $(n_0,m_0)$ are non-negative, i.e., $U_{n_0-1,m_0}(t_0),U_{n_0,m_0-1}(t_0),U_{n_0+1,m_0}(t_0),U_{n_0,m_0+1}(t_0)\geq 0$.  This implies that as $t\rightarrow t_0^+$, the function $U_{n_0,m_0}(t)$ is strictly decreasing. On the other hand, for $t=t_0$, the lattice (\ref{eq1}) becomes:
\begin{eqnarray}
	\label{contr}
	\dot{U}_{n_0,m_0}(t_0)&=&\frac{\gamma_3}{h^2}\Delta_dU_{n,m}(t_0)\nonumber \\&=&\frac{\gamma_3}{h^2}\left(U_{n_0-1,m_0}(t_0)+U_{n_0,m_0-1}(t_0)+U_{n_0+1,m_0}(t_0)+U_{n_0,m_0+1}(t_0)\right)\geq 0,
\end{eqnarray}
because the nonlinear therms of \eqref{eq1} vanish at $t=t_0$ and the neighboring nodes to $U_{n_0,m_0}(t_0)$ are non-negative. Hence, \eqref{contr} implies that $\dot{U}_{n_0,m_0}(t_0)\geq 0$, and consequently, at $t=t_0$, the function $U_{n_0,m_0}(t)$ is increasing or a constant. This conclusion contradicts that at $t\rightarrow t_0^+$ the function $U_{n_0,m_0}(t_0)$ is strictly decreasing.
\end{proof}

By the definition of the discrete Laplacian and the implementation of either case of boundary conditions, we may also prove the following lemma.
\begin{lemma} \label{SDU_0}
Consider the discrete Laplacian supplemented with either case of boundary conditions (vanishing, periodic  
	or  Dirichlet). Then, $\sum_{n,m\in \mathbb{Z}}\Delta_dU_{n,m}=0$.
\end{lemma}
\paragraph{Parametric regimes for decay and uniform bounds} 
With the preparations provided above, we may proceed to the statement of one of the main results concerning the potential extinction scenario described by the decay of solutions or the scenario of the uniform bounds. These crucially depend on certain parametric regimes.
\begin{theorem}
	\label{ASB}
	\begin{enumerate}
		\item \underline{Decay regimes}. For any initial condition $U_{n,m}(0)\geq 0$, $U_{n,m}(0)\in Z$, we have
$\lim_{t\rightarrow\infty}||U(t)||_{\ell^p}=0$ 
in the following cases:
\begin{enumerate}
	\item For all cases of boundary conditions and  $1\leq p\leq\infty$:  when $\alpha=-\tilde{\alpha}<0$, $\beta=-\tilde{\beta}<0$, and $-\tilde{\beta}<-\tilde{\beta}_{\mathrm{thresh}}$ for some
	$-\tilde{\beta}_{\mathrm{thresh}}(\gamma_1, \gamma_2,h)<0$ (depending only on $\gamma_1$, $\gamma_2$, $h$). 
	\item For all cases of boundary conditions and  $2\leq p\leq\infty$: When $\alpha=-\tilde{\alpha}<0$, $\beta>0$ and $-\tilde{\alpha}<-\tilde{\alpha}_{\mathrm{thresh}}$, for some  $-\tilde{\alpha}_{\mathrm{thresh}}(\gamma_1, \gamma_2,\beta,h)<0$.
    \item  For Dirichlet boundary conditions and  $2\leq p\leq\infty$: when $\alpha>0$, $\beta>0$ and $\gamma_3>\gamma_{3,\mathrm{thresh}}:=\frac{\alpha+\gamma}{\mu_1}>0$, for some  $\gamma(\gamma_1,\gamma_2,\beta,h)>0$. 
    \item For Dirichlet boundary conditions and $2\leq p\leq\infty$ : when $\alpha>0$ $\beta=-\tilde{\beta}<0$ and $\gamma_3>\hat{\gamma}_{3,\mathrm{thresh}}:=\frac{\alpha}{\mu_1}>0$,
    and  $-\tilde{\beta}<-\tilde{\beta}_{\mathrm{thresh}}$, where $-\tilde{\beta}_{\mathrm{thresh}}$ is defined in case (a). 
\end{enumerate}
\item \underline{Uniform boundedness regimes}. For any initial condition $U_{n,m}(0)\geq 0$, we have that there exists some $R>0$ such that 
	$\limsup_{t\rightarrow\infty}||U(t)||_{\ell^p}\leq R$, for all $2\leq p\leq\infty$, in the following cases:
	\begin{enumerate}
		\item For all boundary conditions: when $\alpha>0$ and $-\tilde{\beta}<-\tilde{\beta}_{\mathrm{thresh}}$, where $-\tilde{\beta}_{\mathrm{thresh}}$ is defined in case 1.(a).
		\item Periodic or Dirichlet boundary conditions: for arbitrary $\alpha,\beta\in\mathbb{R}$.
	\end{enumerate}
	\end{enumerate}
	\end{theorem}
Proofs for cases 1.($a$) and 1.($c$) are given in the Appendix \ref{App} (see also relevant comments therein). Due to the importance of the desertification effect, a brief numerical study on the relevant scenarios described in Theorem \ref{ASB} follows.	
\subsection{A brief numerical study on the conditions for extinction (desertification)}
We conclude this section, by presenting a brief numerical study relevant to Theorem \ref{ASB}. We choose the case  1.($c$) of Theorem \ref{ASB}. It is an intrigue case, as it provides conditions for extinction when $\alpha,\beta>0$, that is, when aridity effects are against a strong local reproduction $\beta>0$. The condition $\gamma_3>\gamma_{3, \mathrm{thresh}}$ shows that if the short-range cooperative effect $\gamma_3$ is strong enough, extinction is possible. It is also important to emphasize the physical significance of the conditions for extinction, by comparing the cases 1.($c$) and 1.($d$) of Theorem \ref{ASB}. When $\beta > 0$, a stronger condition $\gamma_3 > \gamma_{3,\mathrm{thresh}} = \frac{\alpha + \gamma}{\mu_1}$ guarantees extinction, than the one of case 1.($d$) which refers to the weak local reproduction regime where $\beta < 0$. In the case 1. ($d$), $\gamma_3 > \hat{\gamma}_{3,\mathrm{thresh}} = \frac{\alpha}{\mu_1}$ is a sufficient condition for extinction.

In the numerical study, we choose the following set of parameters: $\alpha=\beta=0.1$ (the aridity effect equals the strong local reproduction), $\gamma_1=0.125$, $\gamma_2=0.5$ (which describe a mediocre strength of long-range competition) in the discrete regime $h=2$ for $L=20$. We use ``tent''-alike, spatially uniform initial conditions of the form  
\begin{equation}\label{eq:tent}
U_{\mathbf{n}}(0) =U^0= \left\{
	\begin{array}{cc}
		A>0, &\text{for all}\quad \mathbf{n}\in \Omega,\\
		0, &\;\;\text{for all}\quad \mathbf{n}\in \partial\Omega,
	\end{array}
	\right.
\end{equation}
\begin{figure}[th!]
\centering 
\includegraphics[scale=0.65]{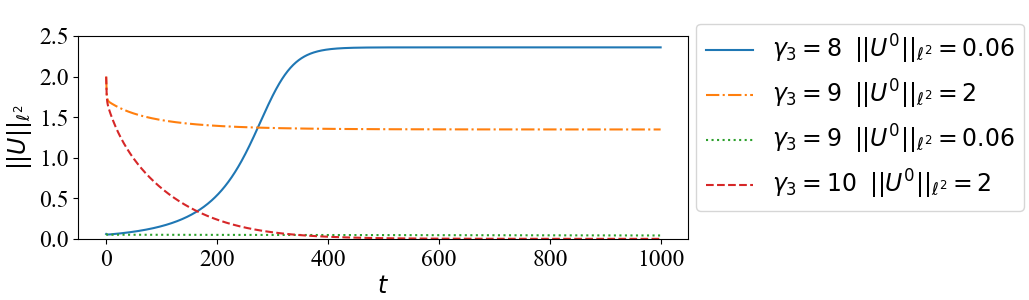} 
\caption{Dynamics of the initial condition \eqref{eq:tent} for various values of $A$, varying the parameter $\gamma_3$. Rest of parameters: $\alpha=\beta=0.1$, $\gamma_1=0.125$, $\gamma_2=0.5$, $h=2$, $L=20$. Blue (solid) curve: $A=0.001$, $\gamma_3=8$. Dotted-Dashed (orange) curve: $A=0.1$, $\gamma_3=9$. Dotted (green) curve: $A=0.001$, $\gamma_3=9$. Dashed (red) curve: $A=0.1$, $\gamma_3=10$. $||U^0||_{\ell^2}$ is the value of the corresponding $\ell^2$-norm of the initial condition \eqref{eq:tent}, for each case of $A$.  Details in the text.}
\label{fig: Extin}
\end{figure}
which satisfy the Dirichlet boundary conditions. The numerical results for their dynamics depicted in Figure 2 highlight the relevance and  high accuracy of the predictions of the theoretical results of Theorem \ref{ASB}-1.($c$). For the selected set of parameters,  the sufficient condition for extinction is $\gamma_3>\gamma_{3, \mathrm{thresh}}>\hat{\gamma}_{3,\mathrm{thresh}} = \frac{\alpha}{\mu_1}=8.1$. The solid (blue) depicts the evolution of the time $\ell^2$-norm of the solution when $A=0.001$ very close to the trivial steady-state. However, since $\gamma_3=8< \hat{\gamma}_{3,\mathrm{thresh}}=8.1$, the dynamics converge to a nontrivial equilibrium, as predicted theoretically.  On the other hand, condition $\gamma_3>  \hat{\gamma}_{3,\mathrm{thresh}}=8.1$ is not sufficient for extinction. This scenario is depicted by the cases for $\gamma_3=9$, where $A=0.1$ [dotted-dashed (orange) curve] and $A=0.001$ [dotted (green) curve], respectively;  still the solutions converge to a nontrivial equilibrium (note that in the case $A=0.001$ the equilibrium has a small norm). For extinction, we need even larger values of $\gamma_3$ as suggested by the sufficient condition $\gamma_3>\gamma_{3, \mathrm{thresh}}>\hat{\gamma}_{3,\mathrm{thresh}}$. Indeed, the numerical experiments detected a numerical threshold $\gamma_{3, \mathrm{num}}$ satisfying $10\gtrapprox\gamma_{3, \mathrm{num}}> \hat{\gamma}_{3,\mathrm{thresh}}=8.1$, for which extinction holds if $\gamma_3>\gamma_{3, \mathrm{num}}$.  This scenario is illustrated by the example of $\gamma_3=10>\gamma_{3, \mathrm{num}}>\hat{\gamma}_{3,\mathrm{thresh}}$ and $A=0.1$ [dashed (red) curve], for which we observe the decaying dynamics of the $l^2$-norm.
 \section{Stability analysis of steady states}
 Theorem \ref{ASB} described the parametric regimes for the tho main dynamical scenarios, decay and potential non-trivial asymptotic behavior (in the sense that solutions may not converge to the trivial steady state). In this section, we investigate further the second scenario, and particularly the physically significant one, of convergence to non-trivial equilibrium. Linear stability analysis will reveal certain criteria on the discretization and the other parameters which highlight the role of discreteness and the dimensionality of the lattice and the differences with the continuous limit (the LL pde).
 \subsection{Linear stability analysis}
The simplest class of equilibria consists of the spatially uniform ones denoted by $U_s$ a constant, that is,  $U_{n,m}=U_s$ for all $n,m\in \mathbb{Z} $, satisfying $f(U_s)=0$. Three distinct branches of spatially uniform steady-states exist:
the trivial branch $U^0_s=0$ and two non-trivial branches
\begin{equation}
\label{two_roots}
U_s^{\pm}=\frac{\beta\pm\sqrt{\beta^2+4\alpha}}{2}\,.
\end{equation}
Notice, that $U^0_s$ exists for all $\alpha,\;\beta$, while $U_s^{\pm}$ exist only when $\beta^2+4\alpha\geq0$.
In particular, 
\begin{itemize}
	\item[(i)] $\beta>0$ implies $U_s^{\pm}>0$ for $\alpha\in \left[-\left(\frac{\beta}{2}\right)^2,0\right)$  and $U_s^{+}>0>U_s^{-}$ for $\alpha\geq 0$. 
	\item[(ii)] $\beta\leq 0$ implies  $U_s^{+}>0>U_s^{-}$ when $\alpha>0$ and $0>U_s^{+}>U_s^{-}$ otherwise.
\end{itemize}
We are interested only on the ecologically realistic equilibria, which must be non-negative, i.e. only the cases of the parametric regimes for $\alpha$ and $\beta$ for which $U_s^{\pm}\geq 0$. Starting with the linear stability analysis, we will denote for simplicity, the non-negative equilibria by $U_s$ and distinguish when appropriate.  For this analysis, we consider perturbations of $U_s$ of the form:
\begin{equation}
\label{eq:pert}
U(t)= U_s +\hat{U}_{n,m}(t),
\end{equation}  
where $ \hat{U}_{n,m}(t)=a(t)e^{i\vec{k}\vec{x}_{n,m}}+cc$ ($cc$=complex conjugates). In \eqref{eq:pert} $\vec{k}=(k_1,k_2)$ is the wavenumber vector of the perturbation and $\vec{x}_{n,m}$ is the vector  $\vec{x}_{n,m}=(nh,mh)$. The function $a(t)$ is the amplitude of the perturbation. The linearized system  around the perturbation \eqref{eq:pert} which is stemming from \eqref{eq1}, is given by the equation
\begin{equation} \label{eq:line_syst}
	\frac{d}{dt} \hat{U}_{n,m} = \mathcal{A} \hat{U}_{n,m}+f'(U_s)  \hat{U}_{n,m}.
\end{equation}
where the linear operator $\mathcal{A}$ is given by
 $$\mathcal{A} =-U_s\frac{\gamma_1}{h^4}\Delta_d^2- (\gamma_2U_s-\gamma_3)\frac{1}{h^2}\Delta_d\,.$$
Actually, by direct substitution of the perturbed solution \eqref{eq:pert} to the linearization (\ref{eq:line_syst}) we derive the following linear ODE for $a(t)$:  
\begin{eqnarray}
\label{parA}
\frac{da(t)}{dt}&=&a(t)\left[-2\hat{U}\frac{(\gamma_2U_s-\gamma_3)}{h^2}\left(\cos(k_1h)+\cos\left(k_2h\right)-2\right)-4\frac{\gamma_1}{h^4}\hat{U}\left(\cos (k_1h )+\cos (k_2h )-2\right)^2+f'(U_s)\right].\;\;\;\;
\end{eqnarray}

 Assuming exponential growth for  $a(t)=a(0)e^{\lambda t}$,  we get from  (\ref{parA}) the eigenvalue problem
\[\left(\mathcal{A}+f'(U_s)\right)a(t)=\lambda a(t),\]
with eigenvalues
\begin{equation}
\label{eq:eigenvalues}
\lambda(k_1,k_2)=f'(U_s) - (\gamma_2U_s-\gamma_3) \frac{2\left[(\cos(k_1h)-1)+(\cos(k_2h)-1)\right]}{h^2}-\gamma_1U_s \frac{4\left[(\cos(k_1h)-1)+(\cos(k_2h)-1)\right]^2}{h^4}\,.
\end{equation}
If $\lambda(k_1,k_2)>0$, $U_s$ is linearly unstable and if $\lambda(k_1,k_2)<0$, $U_s$ is linearly stable.
\paragraph{Linear stability analysis of the trivial steady state.}
As highlighted in Section 3, the trivial steady-state $U^0_s=0$ is of crucial importance for the system since it is associated with desertification, and thus, its stability analysis. For $U^0_s=0$, the eigenvalues are:
\begin{equation}\label{eq:U0_eigen}
\lambda_k^0 = \alpha-\gamma_3\frac{4}{h^2}\left(\sin^2\left(\frac{k_1 h}{2}\right)+\sin^2\left(\frac{k_2 h}{2}\right)\right).
\end{equation}
Therefore, if $\alpha<0$, the  trivial steady-state $U^0_s=0$ is linearly stable for the local dynamics induced by $f$, and it remains stable for non- spatially uniform perturbations, since $\lambda_k^0<0$ for all $k\in \mathbb{R}$.  On the other hand, when $\alpha>0$,  the trivial steady-state is unstable for the uncoupled system. However, large enough values of $\gamma_3$ may linearly stabilize the trivial state, when coupling is present. Particularly, in the case of the  Dirichlet boundary conditions we have the following result.
\begin{proposition}
	\label{stabt}
	Consider the lattice \eqref{eq1} supplemented with the Dirichlet boundary conditions and $\alpha>0$. The steady-state $U_s^0=0$ is linearly stable if $\gamma_3>\dfrac{\alpha}{\mu_1}=\hat{\gamma}_{3,\mathrm{thresh}}$, where $\mu_1$ is the first eigenvalue of the discrete Laplacian.
\end{proposition}
\begin{proof}
In the case of $U^0_s=0$, the linearized system $(\ref{eq:line_syst})$ is simplified to
\begin{equation}
\label{prop4.1}
\frac{d}{dt}\hat{U}_{n,m}=\frac{\gamma_3}{h^2}\Delta_d\hat U_{n,m}+a\hat{U}_{n,m}.
\end{equation}
For perturbations of the form
\begin{equation*}
\hat{U}_{n,m}=b_{n,m}e^{\lambda t}\mbox{, where }\lambda=\lambda (\vec{k}),
\end{equation*}
we derive the eigenvalue problem 
\begin{eqnarray*}
-\frac{\gamma_3}{h^2}\Delta_db_{n,m}=(\alpha-\lambda)b_{n,m},
\end{eqnarray*}
supplemented with Dirichlet boundary conditions. Using the principal eigenvalue $\mu_1$ of the operator $-\Delta_d$ given in \eqref{prince} we deduce that if $\gamma_3>\frac{\alpha}{\mu_1}$, then $U^0_s=0$ is linearly stable.
\end{proof}

It is interesting to remark that in Proposition \ref{stabt} we identified again the threshold value $\hat{\gamma}_{3,\mathrm{thresh}}$ on the parameter $\gamma_3$, this time for the linear stability of $U^0_s=0$, {\em as it was found in Theorem \ref{ASB} 1.(d) for its global stability (i.e., for all initial data) in the case $\beta<0$}. This identification highlights the physical relevance of Theorem \ref{ASB} 1.(d) and Proposition \ref{stabt} concerning the desertification state associated to $U^0_s=0$, with respect to the parameter $\beta$: While the linear stability of $U^0_s=0$ is guaranteed under the condition $\gamma_3>\hat{\gamma}_{3,\mathrm{thresh}}$ for indefinite sign of $\beta$, the same criterion guarantees that $U^0_s=0$ is the global attractor in the case $\beta<0$. On the other hand, the case 1. ($c$) of Theorem \ref{ASB} provides the sufficient condition $\gamma_3>\gamma_{3,\mathrm{thresh}}=\frac{\alpha+\gamma}{\mu_1}$, so that $U^0_s=0$ is the globally attracting state in the case $\beta>0$ (recall the numerical study of section 3.1).
\paragraph{Linear stability analysis of the non-trivial steady-state.}  The results on the instability of the nontrivial spatially homogeneous states are stated in the following Theorem.
\begin{theorem}
	\label{th41}
	Consider the positive steady state $U_s$ and assume that  the parameters of the lattice (\ref{eq1}) satisfy the conditions 
\begin{eqnarray}\label{Condition2}
	C_1&:=&(\gamma_2 U_s-\gamma_3)^2+4\gamma_1U_sf'(U_s)>0,\\
	\label{Condition1}
	C_2&:=&\gamma_3-\gamma_2U_s<0.
\end{eqnarray}	
Then, for every $h$ satisfying 
\begin{equation}
	\label{h_crit}
	h<2\sqrt{\frac{C_2- \sqrt{C_1}}{f'(U_s)}}:=h_c,
\end{equation}
there exist a union $J$ of periodic bands $J_{i,j}$ with empty intersection, such that $\lambda (k_1,k_2)>0$ for all $(k_1,k_2)\in J$ i.e., $U_s$ is linearly unstable.
\end{theorem}
\begin{proof} We start with the simplest case of the absence of linear and nonlinear coupling terms ($\gamma_i=0$).  In this case,  we may easily deduce that the state $U_s>0$ is stable since, by using \eqref{two_roots}, we find that $f'(U_s)<0$.

In the presence of the coupling terms, we will identify the instability bands for wavenumber vector $(k_1, k_2)$, that is, the bands for which the eigenvalues $\lambda(k_1,k_2)$ given in \eqref{eq:eigenvalues}  become positive, and in turns,  $U_s$  loses its stability. {\em A crucial difference with the continuous limit-the Lefever-Lejeune pde-is the presence of the discretization parameter in the eigenvalues \eqref{eq:eigenvalues} which enriches considerably the potential scenarios for instability}, as it will be shown below.

For the analysis, it is convenient to consider the function $\lambda(k_1,k_2)$ \eqref{eq:eigenvalues}, as a composition between the polynomial  
\begin{equation}
\label{polyn}
\Lambda(x)=-\frac{4\gamma_1U_s}{h^4}x^2+2\frac{\gamma_3-\gamma_2U_s}{h^2}x+f'(U_s),
\end{equation}
and the periodic function
\begin{equation}
	\label{pfk}
x(k_1,k_2)=\cos(k_1h)+\cos(k_2h)-2 \in[-4,0],
\end{equation}
with the fundamental period $\left(\frac{2\pi}{h},\frac{2\pi}{h}\right)$.
\begin{figure}[tb!]
	\includegraphics[scale=0.5]{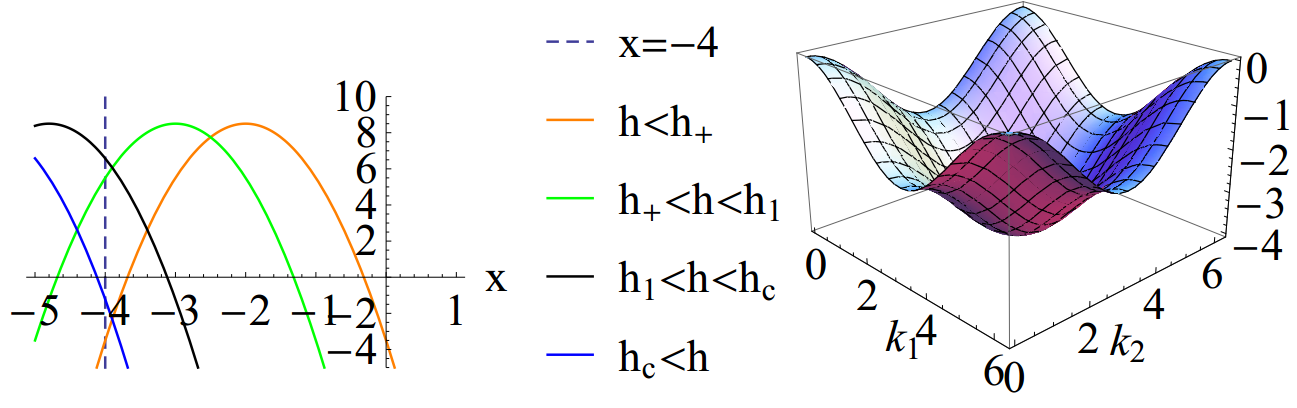}
	\caption{Left panel: Graphs of the polynomial $\Lambda(x)$ \eqref{polyn}, varying $h$ for the $\gamma_1=0.125$, $\gamma_2=0.5$, $\gamma_3=0.5$. $\alpha=0.02$ and $\beta=0.1$. 
Right panel: The graph of the function $x(k_1,k_2)$ \eqref{pfk} in the fundamental period $\left(\frac{2\pi}{h},\frac{2\pi}{h}\right)$, for $h=1$.}
	\label{fig: beta>0}
\end{figure}
Thus, for instability to occur, we need first to ensure the existence of positive range for the polynomial $\Lambda (x)$ \eqref{polyn}:  The  coefficient of the quadratic term of the polynomial $\Lambda(x)$ is negative, so  $\Lambda(x)>0$ for every $x$  between its roots, provided they exist. Thus, we require the discriminant $C_1$ of $\Lambda (x)$ to be positive, which is condition \eqref{Condition2}. Since $f'(U_s)<0$ we see from $\Lambda (x)$ that the product of the roots is positive, thus, the roots have the same sign. In addition, they will be negative if  their sum is negative, which is condition \eqref{Condition1}.

The two roots of the polynomial $\Lambda(x)$ are:
\begin{equation}
	\label{roots}
 r_1=\frac{h^2(C_2-\sqrt{C1})}{4\gamma_1U_s}\;\;\mbox{and}\;\;  r_2=\frac{h^2(C_2+\sqrt{C1})}{4\gamma_1U_s}.
\end{equation}
With conditions (\ref{Condition2}) and (\ref{Condition1}) we ensured that $r_1<r_2<0$. As the second and final step to confirm the instability,  it remains to ensure that the intersection of the range $[-4,0]$ of the function $x(k_1,k_2)$ and of the interval $(r_1,r_2)$ is not empty. The necessary condition is $-4<r_2$ and  this requirement implies, after some algebra, the  criterion \eqref{h_crit}  for the instability of $U_s$ with respect to the discretization parameter $h$. 

Actually, with conditions (\ref{Condition2})-(\ref{Condition1}) we exclude the first graph (blue line) of $\Lambda(x)$ shown in the first panel of Figure \ref{fig: beta>0}.  In the sequel, we will discuss the structure of the instability bands with respect to $h$.  It is useful to recall that the composition of two functions with the same monotonicity is increasing while the composition of two functions with different monotonicity is decreasing.

{\em Case 1: the point  $x_0$ at which $\Lambda(x)$ attains its maximum is included in the range of $x(k_1, k_2)$.}  We start with the case of the graph of $\Lambda(x)$ where the point  $x_0$ at which $\Lambda(x)$ attains its maximum is included in the interval $[-4,0]$ (third (green) and fourth (orange) curves of the first panel of Figure \ref{fig: beta>0}), that is, when the point $x_0$ is included in the range of the function $x(k_1,k_2)$.  We denote for simplicity by $(T_{j_1},T_{j_2})$ any period $\left(\frac{2j_1\pi}{h},\frac{2j_2\pi}{h}\right)$, $j_1,j_2\in\mathbb{N}$, of the function $x(k_1,k_2)$.  We also denote by $\left(k_{j_1}^{(C)},k_{j_2}^{(C)}\right)$ the solution of the equation $x(k_1,k_2)=x_0$  in the sub-domains $(k_1,k_2)\in\left[T_{j_1}, T_{j_1}+\frac{2\pi}{h}\right]\times \left[T_{j_2}, T_{j_2}+\frac{2\pi}{h}\right]$, stating yet for simplicity that is the solution in the $(T_{j_1},T_{j_2})$-period. With some calculus we can verify that the function $x(k_1,k_2)$ has a  minimum in each $(T_{j_1},T_{j_2})$-period  at the point
\begin{eqnarray}
	\label{maximum}
\left(k_{j_1}^{(S)},k_{j_2}^{(S)}\right)=\left(\frac{\pi (2j_1+1)}{h}, \frac{\pi (2j_2+1)}{h}\right).
\end{eqnarray}
The graph of $x(k_1,k_2)$ in the domain defined by the fundamental period $(T_{1},T_{1})$ is plotted in the right panel of Figure \ref{fig: beta>0}.
Therefore, in each $(T_{j_1},T_{j_2})$-period the function $\lambda(k_1,k_2)$ has a  minimum at $(k_{j_1}^{(S)},k_{j_2}^{(S)})$ and a  maximum along the closed level curve  $\cos(hk_1)+\cos(hk_2)-2=x_0$ of $x(k_1,k_2)$, centered at $(k_{j_1}^{(S)},k_{j_2}^{(S)})$. An example of the graph of $\lambda(k_1,k_2)$ depicting the above scenario is shown in the panels of the second row of Figure \ref{fig: Lin_stab}. We also remark  that in the case where $x_0\in(0,4)$, the behavior and graph of  $\lambda(k_1,k_2)$ is similar in either case where one or both of the roots $r_1$ and $r_2$ belong to the interval  $(-4,0)$.  The two negative roots $r_1, r_2$ belong to interval $(-4,0)$ if \eqref{h_crit} holds and  $r_1>-4$, that is, if
\begin{eqnarray}
	\label{h_crit2}
	h<h_+:=2\sqrt{\frac{C_2+\sqrt{C_1}}{f'(U_s)}}.
\end{eqnarray}
It is evident that the larger the distance between the roots $r_1, r_2\in (0,-4)$ is, the more extensive the instability bands become in the  $(k_1,k_2)$-plane.  Another interesting observation which comes out from the formulas of roots \eqref{roots} is the following: Letting $h\rightarrow 0$, which is the formal continuous limit (LL- PDE), $r_{1,2}\rightarrow 0$, and in this limit  $\lambda_c(k_1,k_2)=f'(U_s)-C_2(k_1^2+k_2^2)-\gamma_1U_s (k_1^4+k_2^4)$, which is exactly the characteristic polynomial which is derived  when the linear stability analysis for the continuous system \eqref{eq5} is carried out.  Thus, in the continuous limit, the instability bands shrink at each period, while the periods themselves become infinite.  This scenario is visualized  in the first panel of the first row of Figure \ref{fig: Lin_stab}. Therefore, analytically, it is justified that in the discrete regime, we expect a greater potential for instability of the spatially uniform equilibria.
 
{\em Case 2: the point  $x_0$ at which $\Lambda(x)$ attains its maximum is not included in the range of $x(k_1, k_2)$.}  This is the possibility where the function $\lambda(k_1,k_2)$ has only one maximum in each period $(T_{j_1},T_{j_2})$ (second  (black)  curve of the first panel of Figure \ref{fig: beta>0})). In this scenario, $x_0$ can be identified easily: For $x_0$, we have
\begin{equation*}
\Lambda'(x_0)=0\Leftrightarrow x_0=\frac{C_2h^2}{4\gamma_1U_s}.
\end{equation*}
The polynomial $\Lambda(x)$ is strictly decreasing for $x>x_0$ and therefore, for every $x\in [-4,0]$. Thus, $\lambda(k_1,k_2)$ has one maximum in the interval $(-4,0)$ in each period, if $x_0\notin [-4,0]$, which means that $x_0<-4$. In this case we find that 
\begin{eqnarray}
	\label{h_crit3}
\sqrt{\frac{16\gamma_1U_s}{-C_2}}:=h_1<h.
\end{eqnarray}
Yet with some algebra, we can verify that 
\begin{eqnarray}
	\label{h_crit4}
	h_+<h_1<h_c,
\end{eqnarray}
as expected. In particular, When $h_1<h<h_c$, the function $\lambda(k_1,k_2)$ has only one maximum in the period $(T_{j_1},T_{j_2})$ located at $(k_{j_1}^{(S)},k_{j_2}^{(S)})$ which is the center of this period, see (\ref{maximum}). Therefore, when  $h$ is crossing the value $h_1$ from above, the positions of maximum of the function $\lambda(k_1,k_2)$ tend to coincide and all the points  $(k_{j_1}^{(C)},k_{j_2}^{(C)})$ converge to the point  $(k_{j_1}^{(S)},k_{j_2}^{(S)})$. In addition  every instability band $J_{i,j}$  has as center the point $(k_{j_1}^{(S)},k_{j_2}^{(S)})$ and is bounded by the level curve $-2+\cos(hk_1)+\cos(hk_2)=r_2$, where $r_2$ is the larger root of the polynomial $\Lambda(x)$.
\end{proof}

We remark that Proposition \ref{stabt} remains valid in the case of Dirichlet boundary conditions with the only modification that the number of the corresponding instability bands around the center $\left( k_{j_1}^{(S)},k_{j_2}^{(S)}\right)$ is finite.
%
\begin{figure}[th!]
	\includegraphics[scale=0.4]{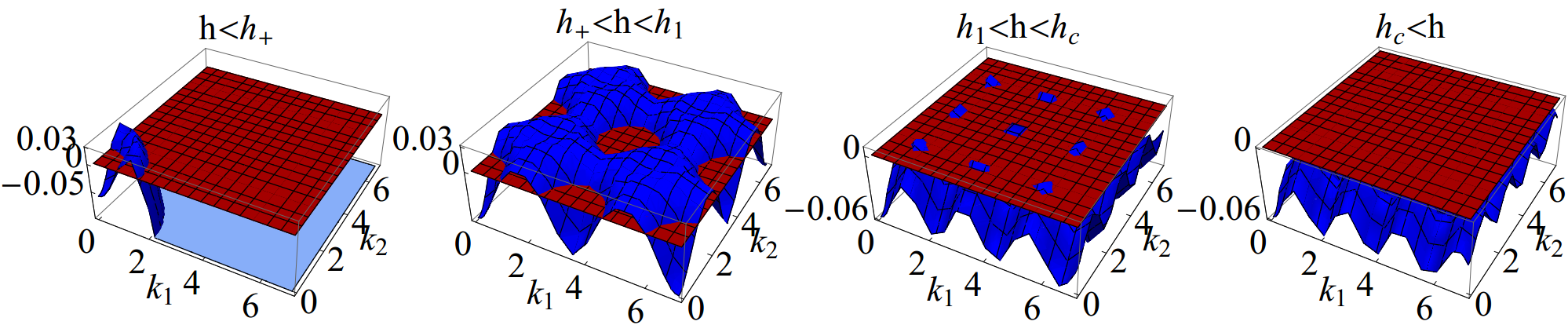}
	\includegraphics[scale=0.4]{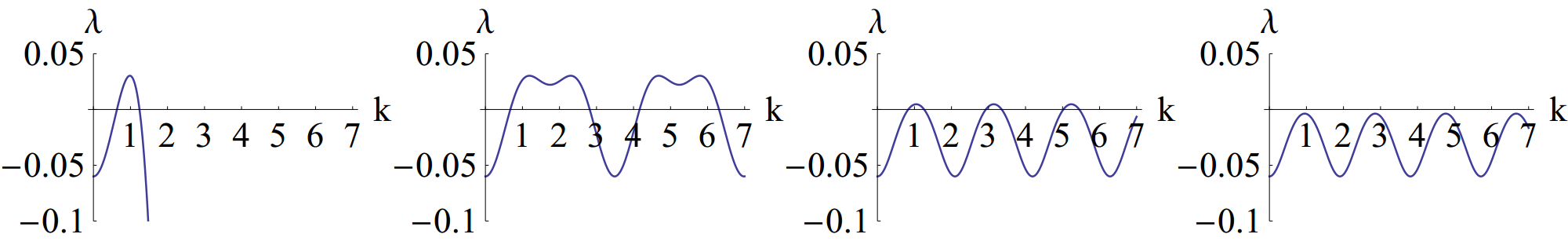}
	\caption{Top row: three-dimensional plots (light (blue) color) of the eigenvalue function $\lambda(k_1,k_2)$ for different  values of the discretization parameter $h$. First panel for $h=0.5\in (0,h_{+})$. Second panel for $h=1.8\in (h_{+},h_1)$. Third panel for $h=3\in (h_1,h_c)$ and fourth panel for $h=3.3>h_c$. The critical values are $h_+=1.63$, $h_1=2.05$ and $h_c=3.16$.  Rest of lattice parameters: $ \gamma_{1}=0.125, \gamma_{2}= 0.5, \gamma_{3}= 0.005, \alpha=0.02, \beta=0.1$.  The dark (red) colored plane is the plane $\lambda=0$.  Bottom row: Each graph depicts the corresponding  cross-sections of the above plots of $\lambda(k_1,k_2)$ for each $h$ along the line $k=k_1$. of the $(k_1,k_2)$-plane. }
\label{fig: Lin_stab}
\end{figure}

Fig.\ref{fig: Lin_stab} visualizes the scenarios of Theorem \ref{th41}, for the set of parameters $ \gamma_{1}=0.125, \gamma_{2}= 0.5, \gamma_{3}= 0.005, \alpha=0.02, \beta=0.1$, depicting the graphs of $\lambda(k_1,k_2)$ when $h$ is varied.

\section{Numerical simulations on the structure of equilibria}\label{section:numerics}
\setcounter{equation}{0}
In this section, we present the results of the numerical simulations for the system \eqref{eq1} supplemented with periodic boundary conditions. We investigate numerically the scenarios of instability of the spatially homogeneous equilibria $U_s$, analyzed in Theorem \ref{th41} and the potential convergence to spatially non-homogeneous states. We distinguish between two cases. The first case studies the dynamics of initial conditions which are spatially periodic perturbations of $U_s$. The second one studies the dynamics of spatially localized initial conditions. 
\subsection{Spatially periodic perturbations of $U_s$: from spatially uniform to spatially periodic equilibria.}
\paragraph{Physical significance and preparations.}
 \label{5para}
 Spatially periodic equilibria which can be considered as perturbations of a spatially uniform state, are relevant to pattern formation of spatially periodic vegetation patches.
These states can be approximated by the simplest periodic ansatz
\begin{eqnarray}
	\label{spans}
	U_P(v_1,v_2)=U_s+A\cos(k_1hv_1)\cos(k_2hv_2),\;\; v_1,v_2 \in \mathbb{Z}.
\end{eqnarray}
We will investigate if such states may appear in the dynamics of the system as a result of the instability of $U_s$ when $\lambda(k_1,k_2)>0$.  In realistic physical scenarios regarding vegetation patterns it is desirable, if spatially periodic structures  can be produced by initial vegetation states which have a similar periodic structure \cite{EhudBook}, by perturbing a spatially uniform vegetation state.  This possibility could be simulated by considering initial conditions of the form 
\begin{equation}\label{eq:IC2}
	U_{n,m}^0=U_{n,m}(0) = U_s +\epsilon \cos\left(k_1hn\right)\cos\left(k_2hm\right),\;\;0<\epsilon<1,\;\;n,m\in\mathbb{Z}.
\end{equation}
If states of the form \eqref{spans} may appear as equilibria of the system for the dynamics initiated from initial conditions \eqref{eq:IC2}, then we could discuss this scenario as an {\em eventual persistence of the spatially periodic patterns}.

With the positive integers $P_1$ and $P_2$ we will denote the periods of the components of discrete function \eqref{spans}, that is $U_P(v_1,v_2)=U_P(v_1+z_1P_1,v_2+z_2P_2)$, where $z_1,z_2\in \mathbb{Z}$.  For the wavenumbers we have the following relations:
Therefore, 
\begin{equation}
	\label{syst_inst}
	\begin{bmatrix}
		&\cos(k_1hv_1)=\cos(k_1h(v_1+P_1)) \\
		&\cos(k_2hv_2)=\cos(k_1h(v_2+P_2))       
	\end{bmatrix}
	\Leftrightarrow
	\begin{bmatrix}
		&k_1hP_1=2\pi n_1 \\
		&k_2hP_2=2\pi n_2 \\
	\end{bmatrix}
	\Leftrightarrow
	\begin{bmatrix}
		&k_1(n,P_1)=\frac{n_1}{P_1}\frac{2\pi}{h} \\
		&k_2(m,P_2)=\frac{n_2}{P_2}\frac{2\pi}{h}
	\end{bmatrix}, \mbox{ where } n_1,n_2\in\mathbb{Z}.
\end{equation}
In other words, the wavenumber vector of the state \eqref{spans} is
\begin{equation}
	\label{kappa1} (k_1,k_2)=\frac{2\pi}{h}\left(\frac{n_1}{P_1},\frac{n_2}{P_2}\right).
\end{equation} 
In the case where  $n_1,n_2\in\mathbb{N}$ and $P_1,P_2\in\mathbb{N}$,  have no mutual factors, the period $(P_1,P_2)$ is called the fundamental period of (\ref{spans}), 
and it is common to call $(n_1,n_2)$ as the ``envelope'', determining the shape of (\ref{spans}).

The relations (\ref{kappa1}) are equivalent to 
\begin{equation}
	\label{KymArith}
	n_i\frac{2 \pi}{k_i}=P_ih\;\;\mbox{for}\;\; i=1,2.
\end{equation}
Therefore, \eqref{KymArith} states that $n_1$ and $n_2$ denote the multiples of the wave lengths in one spatial period for each of the two spatial components. Regarding the potential physical dimension of the wave length of spatially periodic equilibria, we remark that according to the spatial  scaling of (\ref{eq5}), this is given by multiplying $P_1h$ and $P_2h$ by the characteristic inhibition length $L_I$ discussed in Section I.  

\begin{figure}[tbh!]
	\includegraphics[scale=0.6]{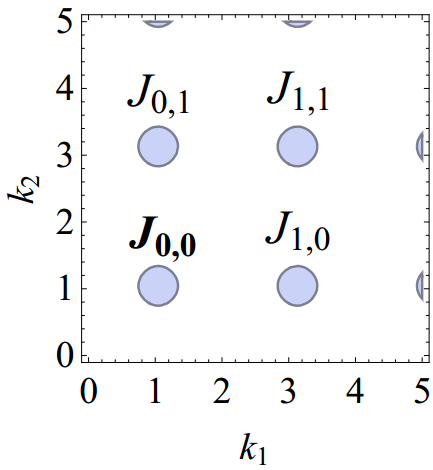}	
	\caption{Instability sets $J_{i,j}$  in the $(k_1,k_2)$-plane when $i=0,1$ and $j=0,1$ for the set of parameters $\alpha=0.02$, $\beta=0.1$,  $\gamma_1=0.125$, $\gamma_2=0.5$, $\gamma_3=0.005$ and $h=3$.}
	\label{fig:3d_disp-relation_Plane}
\end{figure}

With the above preparations, let us implement the instability conditions of Theorem \ref{th41} for $U_s$, in terms the of the initial conditions \eqref{eq:IC2}. The state $U_s$ is unstable if the wavenumber vector \eqref{kappa1}  belongs to an instability set $J_{i,j}$, where $i,j\in \mathbb{N}$. Due to the periodicity of the sets  $J_{i,j}$
it is not a loss of generality to restrict the study in the case where the wavenumber vector \eqref{kappa1} is in the primary instability set $J_{0,0}$. Figure \ref{fig:3d_disp-relation_Plane} depicts the instability sets $J_{i,j}$  in the $(k_1,k_2)$-plane when $i=0,1$ and $j=0,1$. The lattice parameters are $\alpha=0.02$, $\beta=0.1$, $\gamma_1=0.125$, $\gamma_2=0.5$, $\gamma_3=0.005$ and $h=3$. The bounding curve of the set $J_{0,0}$ has equation $-2+\cos(3k_1)+\cos(3k_2)=r_2$, where $r_2=-3.6$ for these parameters.  Figure \ref{fig:3d_disp-relation_Plane} is actually a contour plot of the graph of the function $\lambda(k_1,k_2)$ shown in the third panel of Figure \ref{fig: Lin_stab}. 

For the selection of the above example of parameters we applied the criteria stated in Theorem \ref{th41} as follows: Starting with the branches (\ref{two_roots}) of the spatially homogeneous states $U^{\pm}_s$, $U_s=U^{+}_s$, we select $\alpha=0.02$ and $\beta=0.1$ 
(recall case (i) below Eq. (\ref{two_roots}) for $\alpha\geq 0$). This choice of $\alpha$, $\beta$, defines $U_s=0.2$.  
Next,  by conditions (\ref{Condition2})-(\ref{Condition1}), we derive a threshold value on the parameter $\gamma_3$ for the instability of $U_s$,
\begin{eqnarray}
	\label{dest_for_gamma3}
	\gamma_3<U_s\left(\gamma_2-\sqrt{2\gamma_1(2U_s-\beta)}\right):=\gamma_{3,c}.
\end{eqnarray}
For $U_s=0.2$, we choose $\gamma_1=0.125$ and $\gamma_2=0.5$, and these values define the corresponding $\gamma_{3,c}=0.09$. We select $\gamma_3=0.005$ satisfying the threshold condition for instability. Furthermore, with this choice of parameters the critical value on $h$ given in \eqref{h_crit} is $h_c\cong 3.162$. This is another reason for selecting the above set of parameters in order to investigate the potential sharpness of Theorem \ref{th41} for an example of a strongly discrete system close to the critical value $h_c$ by selecting $h=3$. 

For the above example of the lattice \eqref{eq1}, representative wavenumber vectors $(k_1,k_2)\in J_{0,0}$  in the simplest case of $k_1=k_2$ and some fundamental periods $P_i$, are given in the following table.
\begin{center}
\begin{tabular}{|c|c|c|}
\hline 
$n_i$  & $P_i$ & $k_1=k_2$ \\
\hline
1 & 2 & 1.0472 \\
2 or  3 & 5 & 0.8377 or  1.2566\\
3 or  4 & 7 & 0.8975  or  1.1967\\
\hline
\end{tabular}
\end{center}

\begin{figure}[th!]
	\centering 
\includegraphics[scale=0.35]{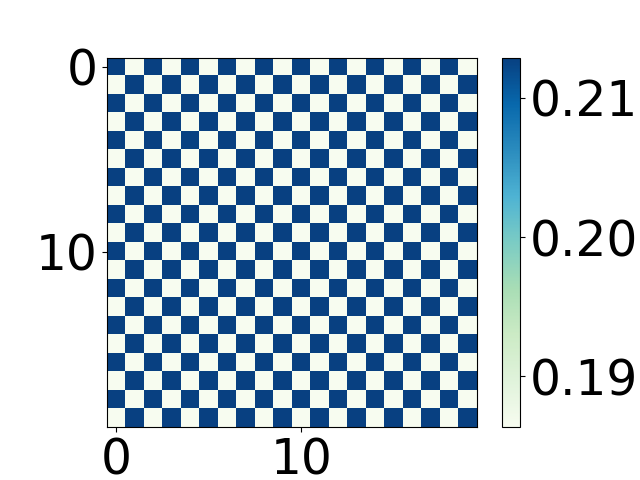} 
\includegraphics[scale=0.35]{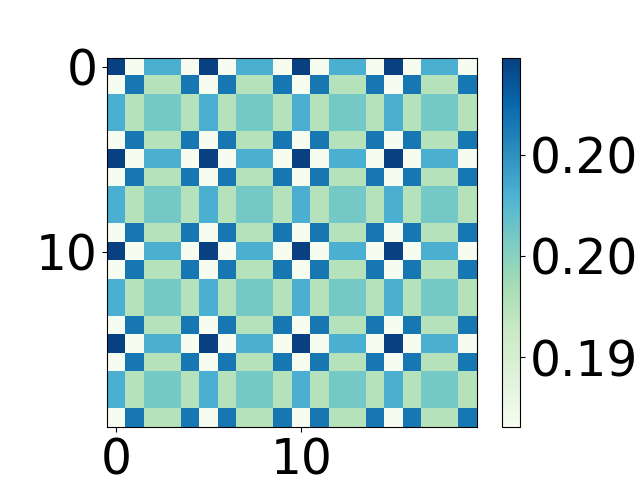}
 \includegraphics[scale=0.35]{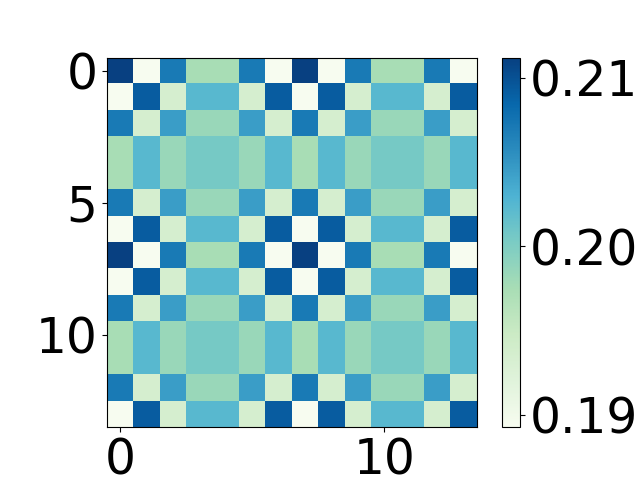}
\caption{
Contour plots of the spatially periodic equilibria to which the dynamics of the initial conditions \eqref{eq:IC2} converge for the lattice parameters $\alpha=0.02$, $\beta=0.1$, $\gamma_1=0.125$, $\gamma_2=0.5$, $\gamma_3=0.005$, $h=3$ (for their selection see paragraph $5.1.a$). Middle panel: $n=m=1$ and $P_1=P_2=2$, $L=30$ for the square lattice $\Omega=[-L,L]\times[-L,L]$.  Middle panel: $n=m=2$ and $P_1=P_2=5$, $L=30$. Right panel: $n=m=3$ and $P_1=P_2=7$, $L=21$. Details in the text.
}
\label{fig: periodics}
\end{figure}

\paragraph{Numerical results for the spatially periodic equilibria.}
Figure \ref{fig: periodics} depicts the contour plots for three cases of the equilibria to which the initial conditions \eqref{eq:IC2} converge,  when their wavenumbers are chosen in the unstable set $J_{0,0}$.  The example  of the set of the parameters of the lattice is the one selected by the process described in paragraph $5.1.a$ above. 
For the initial conditions \eqref{eq:IC2}, $U_s=2$ as dictated by the prescribed set of lattice parameters and $\epsilon=0.01$. 
Also, the corresponding unstable wavenumber vectors, relevant to $n=m=n_1=n_2$ and the periods $P_1=P_2$ are selected from the table of paragraph $5.1.a$.

In all cases, we observe that the equilibrium states are spatially periodic  ``mosaic'' patterns, with the highlight that they preserve the wavenumbers of the initial condition as well as its period; this is evident by examining the period of each node which composes the internal pattern in each equilibrium (see in particular the equilibria of the middle and right panels). This preservation can be explained by the fact, that since they are intially selected from the instability set $J_{0,0}$ they can be the only ones who can ultimately survive due to the instability of $U_s$. It is another justification that the system undergoes a Turing instability mechanism demonstrated in the discrete set-up. Recalling that we have selected $h=3<h_c=3.162$, the numerical results illustrate the sharpness of the instability condition \eqref{h_crit} of Theorem \ref{th41}.
%
\begin{figure}[tbh!]
	\centering 
	\includegraphics[scale=0.82]{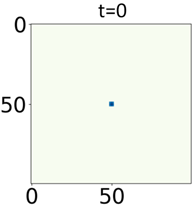} 
	\includegraphics[scale=0.82]{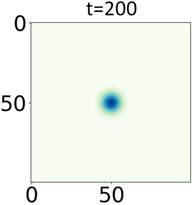}
	\includegraphics[scale=0.82]{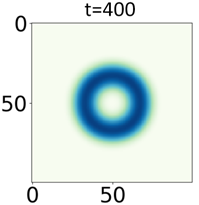}
	\includegraphics[scale=0.82]{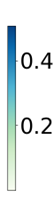}\\
	\includegraphics[scale=0.82]{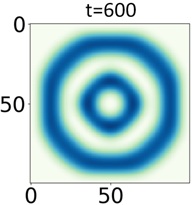} 
	\includegraphics[scale=0.82]{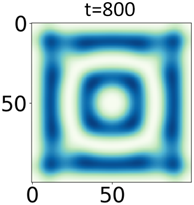}
	\includegraphics[scale=0.82]{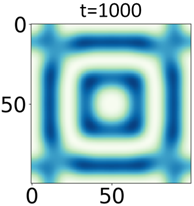}
	\includegraphics[scale=0.82]{Fig6_bar.png}
	\caption{
		Snapshots of the evolution of the initial condition  \eqref{eq:ICS} for $A=0.3$. Lattice parameters: $\alpha=0.05$, $\beta=0.1$, $\gamma_1=0.125$, $\gamma_2=0.5$, $\gamma_3=0.005$, $h=0.2$, $L=10$. Details in the text.
	}
	\label{fig6}
\end{figure}
\begin{figure}[th!]
	\centering 
	\includegraphics[scale=0.69]{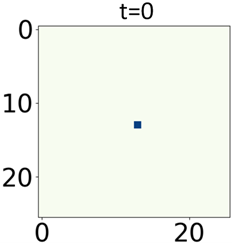} 
	\includegraphics[scale=0.69]{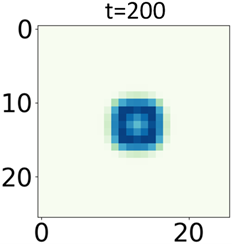}
	\includegraphics[scale=0.69]{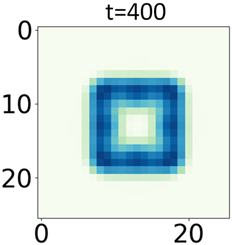}
	\includegraphics[scale=0.69]{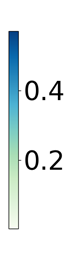}\\
	\includegraphics[scale=0.69]{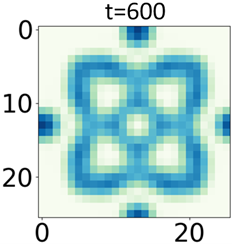} 
	\includegraphics[scale=0.69]{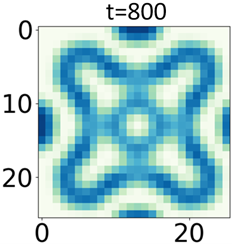}
	\includegraphics[scale=0.69]{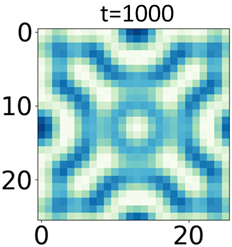}
	\includegraphics[scale=0.69]{Fig7_bar.png}
	\caption{
		Snapshots of the evolution of the initial condition  \eqref{eq:ICS} for $A=0.3$ for the same set of parameters as in Figure \ref{fig6}, but for $h=0.8$, $L=10$. Details in the text.
	}
	\label{fig7}
\end{figure}
\begin{figure}[tbh!]
	\centering 
	\includegraphics[scale=0.38]{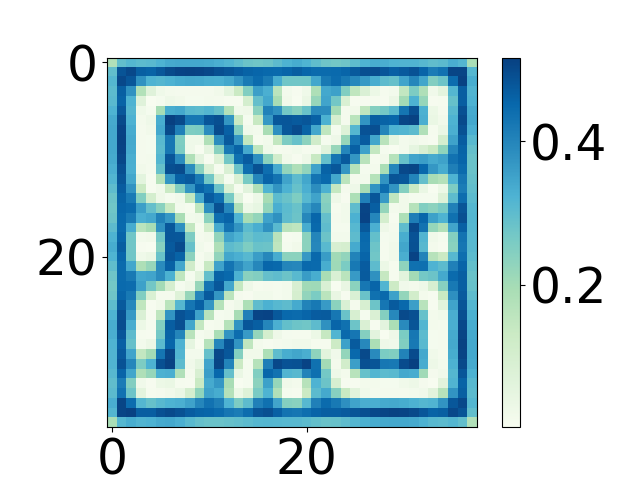} 
	\includegraphics[scale=0.38]{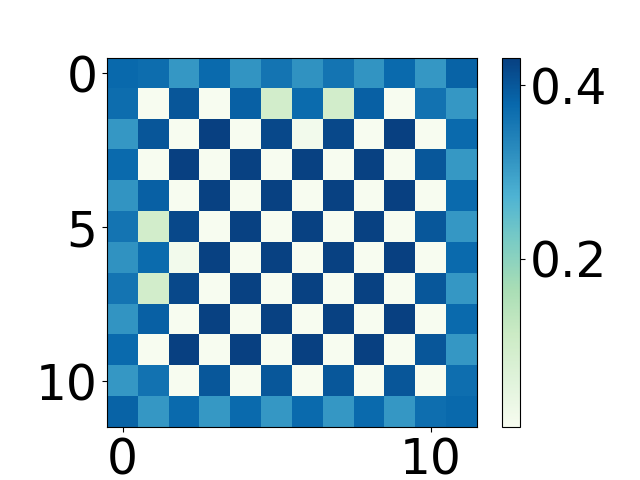}
	\caption{
		Contour plots of the spatially nonuniform equilibria to which the dynamics of the initial condition \eqref{eq:ICS} with $A=0.3$. The lattice parameters are fixed as in Figure \ref{fig6} except of $h$ and $L$. Left panel: $h=0.8$, $L=16$. Right panel: $h=2.5$, $L=15$. 
	}
	\label{fig8}
\end{figure}
\begin{figure}[th!]
	\centering 
	\includegraphics[scale=0.35]{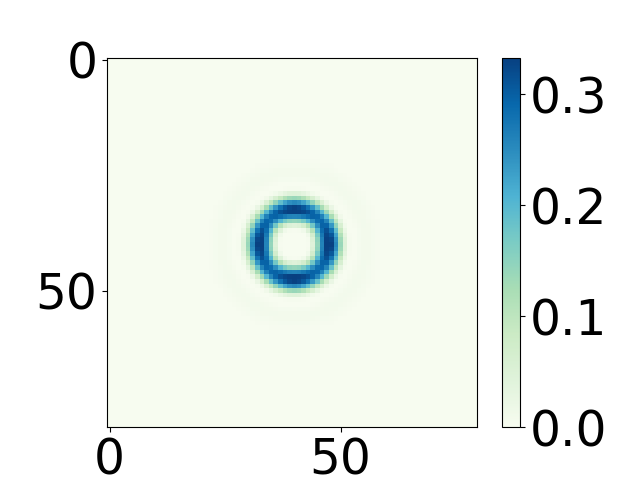} 
	\includegraphics[scale=0.35]{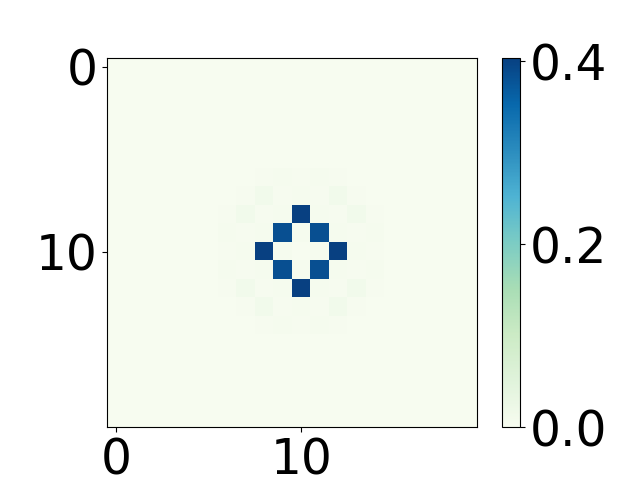}
	\includegraphics[scale=0.35]{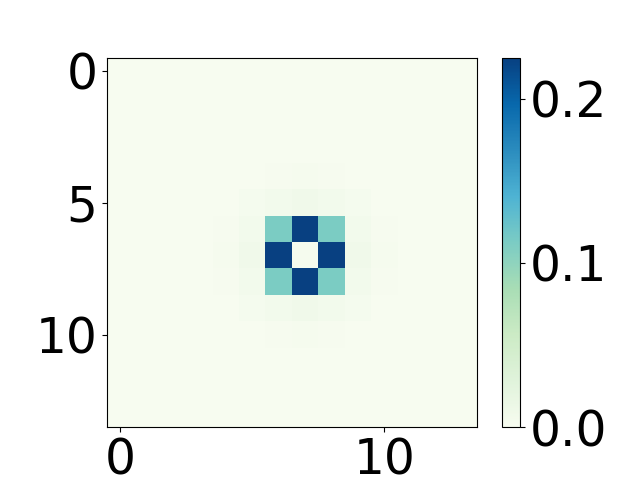}\\
	\includegraphics[scale=0.35]{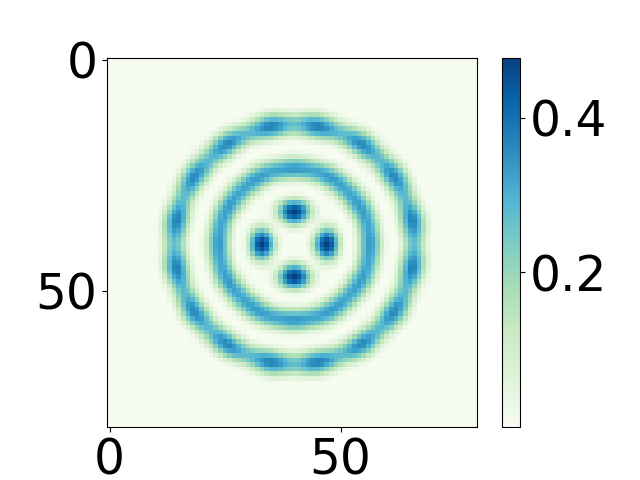}
	\includegraphics[scale=0.35]{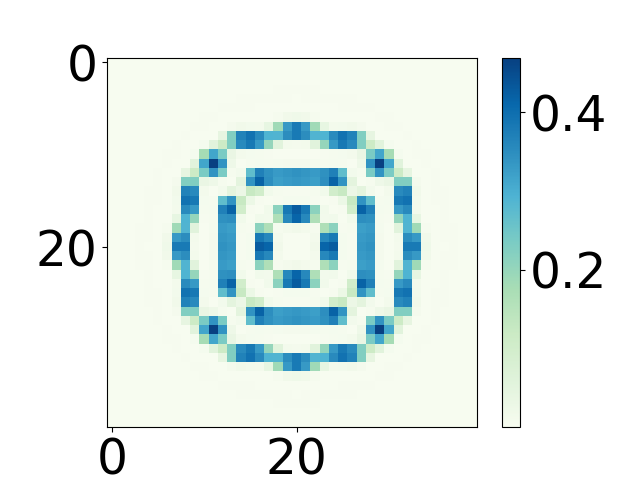} 
	\includegraphics[scale=0.35]{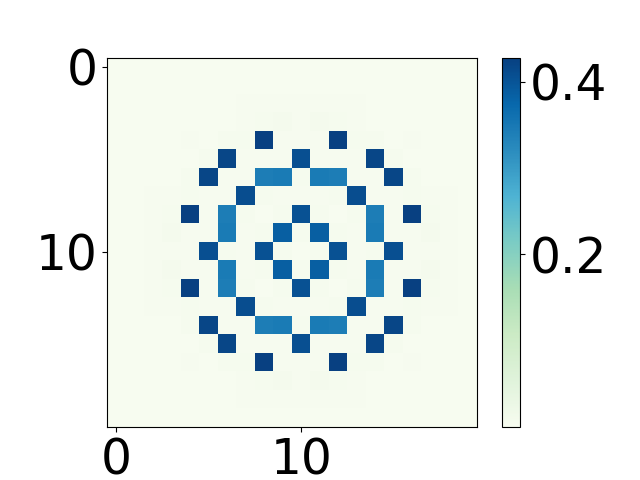}
	\caption{Contour plots of the equilibria of the dynamics of centered a Gaussian intial conditions of amplitude $A=0.3$ and variance $\sigma$. Lattice parameters: $\alpha=-0.002$, $\beta=0.1$, $\gamma_1=0.125$, $\gamma_2=0.5$ and $\gamma_3=0.005$.  Top row for $\sigma=3$. Left panel: $h=0.5$, $L=20$. Middle panel: $h=2$, $L=20$. Right panel: $h=3$, $L=20$.  Bottom row for $\sigma=6$. Left panel: $h=0.5$, $L=20$. Middle panel: $h=1$, $L=20$. Right panel: $h=2$, $L=20$.
	}
	\label{fig9}
\end{figure}
\subsection{Localized initial data: Invasion phenomena and localized states}
 It is interesting to study the dynamics of the lattice  \eqref{eq1} in the case where the destabilization conditions of Theorem \ref{th41} for the nontrivial steady state $U_s$ are valid and the initial conditions are spatially  localized. At this point, it is important to emphasize the following: the instability criteria outlined in Theorem \ref{th41} remain valid even for generic initial conditions. This is attributed to the ability to analyze any initial condition through the (discrete) Fourier transform. The wavenumbers within the Fourier spectrum which are in the instability sets $J_{i,j}$, propel the dynamics toward instability. Conversely, those wavenumbers which are not in  $J_{i,j}$ dissipate, leading to convergence to the spatially uniform state $U_s$. The   outcome is the convergence of the generic initial condition to a spatially non-uniform equilibrium.
 
  The numerical simulations illustrate that the dynamics and the corresponding patterns have a rich structure. We study the dynamics for two cases of initial conditions: ($a$) initial excitation of a single node and ($b$) Gaussian initial conditions. We also consider the two cases of productivity regimes $\alpha>0$ and $\alpha<0$ for small $|\alpha|$.
 \paragraph{Initial excitation of a single node for $\alpha>0$.} 
Figure \ref{fig6} shows snapshots of the evolution of the initial condition
\begin{equation}\label{eq:ICS}
	U_{n,m}(0) = \left\{
	\begin{array}{cc}
		A>0, &\text{for}\quad n=m=0,\\
		0, &\text{for all}\quad n,m\neq 0,
	\end{array}
	\right.
\end{equation}
for $A=0.3$. The parameter $h=0.2$, that is, close to the continuous limit. Since $h<1$ both competition and inhibition mechanisms are significant.  The parameter $\alpha=0.05$, which means that the environment has limited aridity which still ensures productivity; recall from Section 1, that the mortality to growth rate ratio $\mu=1-\alpha$, thus $\mu<1$.  The rest of parameters are  $\beta=0.1$, $\gamma_1=0.125$, $\gamma_2=0.5$ and $\gamma_3=0.005$. For this set of parameters, we find $h_c=2.16$, thus the condition for the instability of $U_s=0.27$ holds. The snapshots illustrate an invasion phenomenon due to high productivity regime. The final state is achieved almost at $t=1000$. 

Increasing $h$, the numerical simulations illustrate that  the invasion dynamics can be very rich as well as, the structure of the equilibrium states. This is demonstrated in Figure \ref{fig7} depicting the evolution of the initial condition \eqref{eq:ICS} for the same set of parameters as above but for $h=0.8$.  The size of the lattice (and thus the number of interacting nodes)  it is natural to affect the form of the pattern. The first panel of Figure \ref{fig8} shows the contour plot of the equilibrium when the discretization parameter  is still fixed to $h=0.8$ but the length is changed to $L=16$. For increased  $h=2.5$ and for $L=15$, the invasion process converges to the "mosaic" pattern shown in the right panel of Figure \ref{fig8}.

\paragraph{Gaussian initial conditions for $\alpha<0$.} We conclude with the presentation of numerical results in the case where $\alpha=-0.002<0$. Such an example means that the environment has an aridity which is leading to low productivity/mortality. We use again the set of parameters $\beta=0.1$, $\gamma_1=0.125$, $\gamma_2=0.5$ and $\gamma_3=0.005$ and we find that $h_c=8.64$. 
With such conditions it is natural to expect localization phenomena than invasion when $h<h_c$ and the uniform state $U_s=0.07$ is unstable (observe that it is close to the trivial state). 

This behavior is illustrated in Figure \ref{fig9}, depicting the contour plots of the equilibria to which the dynamics of Gaussian initial data converge.  The amplitude of the Gaussian initial condition is $A=0.3$ and its expected value is zero. The top row depicts the contour plots of the equilibria when the variance of the initial condition is  $\sigma=3$ and the bottom row when $\sigma=6$. The parameter $h$ varies in each row. For $\sigma=3$, it is $h=0.5$ in the first panel, $h=2$ in the middle panel and $h=3$ in the right panel, for $L=20$.  For $\sigma=6$,  it is $h=0.5$ in the first panel, $h=1$ in the middle panel and $h=2$ in the right panel, for $L=20$.  This selection of equilibria provides another evidence of the rich morphology of the patterns exhibited by the dynamics of the DLL system.
\section{Conclusions} \label{conclusions}
In this paper, we aimed to underscore the significance of spatial discreteness \cite{durrett1994} within the context of ecosystem physics. Our approach centered around the examination of the two-dimensional discrete Lefever-Lejeune equation, derived from the spatial discretization of the renowned Lefever-Lejeune partial differential equation. We posited that, inherently, this model holds relevance for describing vegetation dynamics in drylands. Specifically, we explored the role of discreteness as a measure for the overlap of zones of influence between individual plant dynamics, treating biomass patches akin to interacting particles.

We studied the model as an evolution equation in the appropriate sequence spaces defined by the boundary conditions supplementing the model. We established estimates for both the decay of solutions, representing the physical phenomenon of extinction or desertification, and uniform bounds, suggesting convergence to a non-trivial attractor. Given the two-dimensional nature of the lattice, it serves as an ideal environment for potential pattern formation.

Analytically, we verified the manifestation of a Turing instability mechanism, contingent on explicit thresholds for the discretization parameter. If satisfied, these thresholds guide the dynamics, through the instability of spatially uniform steady states, toward a diverse set of spatially non-uniform equilibria. Consequently, the resulting patterns are diverse, as illustrated by numerical simulations: (a) Periodic 'mosaic' patterns, preserving the structure of periodic perturbations of the uniform states when used as initial conditions. This effect is explained by the survival of unstable modes due to Turing instability.
(b) More complex patterns resulting from invasion dynamics initiated by localized initial conditions in the weak productivity regime. Here, localized structures coexist with periodic ones.
(c) Localized structures emerging from localized initial conditions in the weak mortality regime, demonstrating isolated spots and rings, or combinations thereof.

The above results, apart from emphasizing the crucial role of spatial discreteness in the context of mathematical ecology, motivate further investigations. Of particular interest is the exploration, following the approach of the present paper, of nonlinear lattices such as the Lefever-Lejeune or the discrete Swift-Hohenberg  equation \cite{Pelet1}, \cite{Pelet2}, incorporating spatial forcing \cite{Eforc}. These counterparts are significant, as external forcing terms may effectively model external, even human, interference in vegetation dynamics. Works in this direction are in progress and will be reported elsewhere. 

\vspace{-0.5cm}
\section*{Acknowledgments}
The paper is dedicated to Professor Ioannis G. Stratis, a closed friend, collaborator and mentor whose scientific guidance  is always  a continuous source of inspiration.
\vspace{-0.5cm}
\section*{Funding statement}
There are no funders to report for this submission.
\vspace{-0.5cm}
\section*{Conflict of interest statement}
This work does not have any conflicts of interest.
\vspace{-0.5cm}
\section*{Authors contributions statement}
All authors contributed equally to the study conception, design and writing of the manuscript. Material preparation, data collection and analysis were performed equally by all authors. All authors read and approved the final manuscript.
\appendix
\section{Proof of cases 1.($a$) and 1.($c$) of Theorem \ref{ASB}}
\label{App}
For the sake of completeness, we indicatively provide the proofs for cases 1.($a$) and 1.($c$), since the other cases can be proved by similar arguments, along the lines of \cite{LLPhysD}.  We note that for  the proof of case 1.($a$), we use that solutions are non-negative (which is a new result which was not considered in \cite{LLPhysD}) and is of particular physical significance for the model. Furthermore, it allows to prove the decay of solutions in the strongest $\ell^1$-norm.  On the other hand, the proof of the case 1.($c$) highlights the crucial role of the discrete Poincar\'{e} inequality \eqref{crucequiv}, in order to derive a criterion for the global stability of the trivial steady state in the regime $\alpha,\beta>0$. \newline
{\em Proof of case 1.($a$)}: First note that due to  (\ref{genlp8}) and (\ref{genlp9}), the following three inequalities are valid for all cases of boundary conditions:
	\begin{eqnarray}
		\label{anisot1}
		\left|\sum_{n,m\in\mathbb{Z}}U_{n,m}\Delta_dU_{n,m}\right| &=&\vert\left<U,\Delta_d U\right>_{\ell^2}\vert\leq ||U||_{\ell^2}||\Delta_dU||_{\ell^2} \leq C||U ||_{\ell^2}^2,   \\
		\label{anisot2}
		\left|\sum_{n,m\in\mathbb{Z}}U_{n,m}\Delta_d ^2U_{n,m}\right|&=&|\left<U,\Delta_d^2U\right>_{\ell^2}| \leq || U||_{\ell^2} || \Delta _d^2U||_{\ell^2} \leq \hat{C} || U ||_{\ell^2}^2.
	\end{eqnarray}
	In  (\ref{eq1}), due to Lemma (\ref{posit})  we may omit the term $U_{n,m}(t)^3\geq 0$ for all $t\geq 0$, to get the following inequality for every $n,m\in {\mathbb{Z}}$:
	\begin{eqnarray}
		\label{IN1}
		\dot{U}_{n,m}+\tilde{\alpha} U_{n,m}+\tilde{\beta} U_{n,m}^2+\frac{\gamma_1}{h^4}U_{n,m}\Delta_d^2U_{n,m}+ \frac{\gamma_2}{h^2}U_{n,m}\Delta_dU_{n,m}-\frac{\gamma_3}{h^2}\Delta_dU_{n,m}\leq 0.
	\end{eqnarray}
	Summation in \eqref{IN1} and application of \eqref{anisot1} and \eqref{anisot2} implies the following estimate:
	\begin{eqnarray*}
		\sum_{n,m\in\mathbb{Z}}\dot{U}_{n,m}-\frac{\gamma_3}{h^2}\sum_{n,m\in\mathbb{Z}}\Delta_dU_{n,m}&+&\tilde{\alpha}\sum_{n,m\in\mathbb{Z}}U_{n,m}+\tilde{\beta}\sum_{n,m\in\mathbb{Z}}U_{n,m}^2 \\
		&\leq &\frac{\gamma_1}{h^4}\left|\sum_{n,m\in\mathbb{Z}}U_{n,m}\Delta_d^2U_{n,m}\right|+\frac{\gamma_2}{h^2}\left|\sum_{n,m\in\mathbb{Z}}U_{n,m}\Delta U_{n,m}\right|\\
		&\leq &\frac{\hat{C}\gamma_1}{h^4}|| U||_{l^2}^2+\frac{C\gamma_2}{h^2}|| U||_{\ell^2}^2.
	\end{eqnarray*}
	The above inequality can be rewritten in the form
	\begin{eqnarray}
		\label{eq14}
		\frac{d}{dt}\sum_{n,m\in\mathbb{Z}}U_{n,m}\leq\left(\frac{\hat{C}\gamma_1}{h^4}+\frac{C\gamma_2}{h^2}-\tilde{\beta}\right)|| U||_{l^2}^2-\tilde{\alpha}\sum_{n,m\in\mathbb{Z}}U_{n,m}.
	\end{eqnarray}
	Setting $\tilde{\beta}_{\mathrm{thresh}}(\gamma1,\gamma_2,h)=\frac{\hat{C}\gamma_1}{h^4}+\frac{C\gamma_2}{h^2}$, we deduce that for all  $\tilde{\beta}>\tilde{\beta}_{\mathrm{thresh}}$ 
	\begin{eqnarray*}
		\label{AniParag}
		\frac{d}{dt}\sum_{n,m\in\mathbb{Z}}U_{n,m}+\tilde{\alpha}\sum_{n,m\in\mathbb{Z}}U_{n,m}\leq 0,
	\end{eqnarray*}
	and the standard Gronwall lemma implies decay of the $\ell^1$-norm, since $U_{n,m}(t)\geq 0$ for all $t>0$ and
	\begin{equation}
		\label{lim_0}
		\lim_{t\rightarrow +\infty} \sum_{n,m\in\mathbb{Z}}U_{n,m}=0.
	\end{equation} 
	From \eqref{eq7imb}, $||U||_{\ell^p}\leq ||U||_{\ell^1}$, for all $1\leq p\leq\infty$, thus \eqref{lim_0} implies the decay in all  $\ell^p$-norms.
	
	{\em Proof of case 1.($c$)}: For  $\alpha$ and $\beta>0$, after multiplication of Eq. (\ref{eq1}) with $U$ in the $\ell^2$-inner product, we get the balance equation,
\begin{eqnarray}
	\label{eq1c1}
	\frac{1}{2}\frac{d}{dt}||U||^2_{\ell^2}-\frac{\gamma_3}{h^2}\left<\Delta_dU, U\right>_{\ell^2}+||U||_{\ell^4}^4 -\alpha||U||^2_{\ell^2}=-\frac{\gamma_1}{h^4}\left<U\Delta^2_dU, U\right>_{\ell^2}-\frac{\gamma_2}{h^2}\left<U\Delta_dU, U\right>_{\ell^2}+\beta\left<U^2, U\right>_{\ell^2}.
\end{eqnarray}	
By using inequalities \eqref{genlp8} and \eqref{genlp9}, each term in the right-hand side of \eqref{eq1c1}, is estimated as follows:
	\begin{eqnarray}
		\label{eq11n}
		\frac{\gamma_1}{h^4}\left|\left<U\Delta^2_dU, U\right>_{\ell^2}\right|&\leq&\frac{\gamma_1}{h^4}||U^2||_{\ell^2}||\Delta^2_dU||_{\ell^2}=\frac{\gamma_1}{h^4}||U||_{\ell^4}^2||\Delta^2_dU||_{\ell^2}\leq \frac{\hat{C}\gamma_1}{h^4}||U||_{\ell^4}^2||U||_{\ell^2}\nonumber\\
		&\leq &  \frac{1}{\varrho^2}||U||_{\ell^4}^4+\frac{\hat{C}^2\gamma_1^2\varrho^2}{ h^8}||U||_{\ell^2}^2
		= \frac{1}{\varrho^2}||U||_{\ell^4}^4+c_1||U||_{\ell^2}^2,\\ 
		&\textit{ where }& c_1=c_1(\gamma_1,h,\varrho)=\left(\frac{\hat{C}\varrho \gamma_1}{h^4}\right)^2, \nonumber \\
		\label{eq12n}
		\frac{\gamma_2}{h^2} \left|\left<U\Delta_dU, U\right>_{\ell^2}\right|&\leq &\frac{\gamma_2}{h^2}||U^2||_{\ell^2}||\Delta_dU||_{\ell^2}=\frac{\gamma_2^2}{h^2}||U||_{\ell^4}^2||\Delta_dU||_{\ell^2}= \frac{C\gamma_2}{h^2}||U||_{\ell^4}^2||U||_{\ell^2}\nonumber \\
		&\leq& \frac{1}{\varrho^2}||U||_{\ell^2}^4+\frac{C^2\varrho^2\gamma_2^2}{h^4}||U||_{\ell^2}^2  = \frac{1}{\varrho^2}||U||_{\ell^4}^4+c_2||U||_{\ell^2}^2, \\	
		&\textit{ where }& c_2=c_2(\gamma_2,h,\varrho)=\left(\frac{C\varrho\gamma_2}{h^2}\right)^2, \nonumber \\
		\label{eq13n}
		\beta\left|\left<U^2, U\right>_{\ell^2}\right|&\leq&\beta||U^2||_{\ell^2}||U||_{\ell^2}=\beta||U||_{\ell^4}^2||U||_{\ell^2}
		\leq \frac{1}{\varrho^2}||U||_{\ell^4}^4+\beta^2\varrho^2||U||_{\ell^2}^2\nonumber \\
		&=&\frac{1}{\varrho^2}||U||_{\ell^4}^4+c_3||U||_{\ell^2}^2,\\
		&\textit{ where }& c_3=c_3(\beta,\varrho)=\beta^2\varrho^2.\nonumber
	\end{eqnarray}	
	Note that for the estimates \eqref{eq11n}-\eqref{eq13n},  we have used Young's inequality $ab\leq \frac{a^2}{\varrho^2}+ b^2\varrho^2$, for $a,b\geq 0$. The second term of the left-hand side of \eqref{eq1c1} is estimated from below by using the {\em discrete Poincar\'{e} inequality \eqref{crucequiv}}:
\begin{equation}
	\label{eq1cpi}
	\gamma_3\mu_1||U||_{\ell^2}^2\leq
	\frac{\gamma_3}{h^2}\left<-\Delta_dU,U\right>_{\ell^2}.
\end{equation}	
Inserting the estimates \eqref{eq11n}-\eqref{eq13n} and \eqref{eq1cpi} into the balance equation \eqref{eq1c1}, we derive the existence of a constant $\gamma(c_1, c_2, c_3)=c_1+c_2+c_3>0$, such that 
\begin{eqnarray}
	\label{eq24nn}
	\frac{1}{2}\frac{d}{dt}||U||^2_{\ell^2}+\left(\gamma_3\mu_1-\alpha-\gamma\right)||U||_{\ell^2}^2+\left (1-\frac{3}{\varrho^2}\right)||U||_{\ell^4}^4<0.
\end{eqnarray}
Selecting $\varrho>\sqrt{3}$, we get that if $\gamma_3>\gamma_{3,\mathrm{thresh}}=(\alpha+\gamma)/\mu_1$,the following estimate holds
\begin{eqnarray*}
	||U(t)||^2_{\ell^2}\leq \mathrm{e}^{-2\Gamma t}||U(t)||^2_{\ell^2},
\end{eqnarray*}	
with $\Gamma=\gamma_3\mu_1-\alpha-\gamma>0$, implying the decay of solution as $t\rightarrow\infty$. \ \ $\Box$
\section{Continuity properties of discrete linear operators.}
\label{AppB}
Herein, yet for the sake of completeness, we provide the proofs of the inequalities \eqref{genlp8} and \eqref{genlp9} due to their importance for the derivation of various estimates (see Theorem \ref{ASB}), particularly in the cases of the finite lattices.

The discrete Laplacian in $\mathbb{Z}^2$ can be written in the form 
\begin{eqnarray}
\label{DefLap_n}
\Delta_d=\Delta_{(dn)}U_{n,m}+\Delta_{(dm)}U_{n,m},
\end{eqnarray}
where $\Delta_{(dn)}U_{n,m}=U_{n+1,m}-2U_{n,m}+U_{n-1,m}$ and $\Delta_{(dm)}U_{n,m}=U_{n,m+1}-2U_{n,m}+U_{n,m-1}$,
are the discrete Laplacian operators considered in the $n$- direction and $m$-direction of the $\mathbb{Z}^2$-lattice, respectively. 
Then, the discrete biharmonic operator 
$\Delta^2_d=\Delta_d(\Delta_d)$ has the form,
\begin{eqnarray}
\label{BDOp}
\Delta^2_dU_{n,m}=\Delta_d(\Delta_dU_{n,m})=\Delta^2_{(dn)}U_{n,m}+\Delta_{(dn)}(\Delta_{(dm)}U_{n,m})+\Delta_{(dm)}(\Delta_{(dn)}U_{n,m})+\Delta^2_{(dm)}U_{n,m}.
\end{eqnarray} 
Inequality \eqref{genlp8} is proved in the following lemma.
\begin{lemma}
\label{LB1}
Let $Z={\ell^p_{\mathrm{per}}}, \ell^2_0$. The discrete differential operator $\Delta_d :Z\rightarrow \ell^p$ satisfies the inequality $\|\Delta U\|_{\ell^p}\leq C\|U\|_Z$, for all $U\in Z$.
\end{lemma}
\begin{proof} We will give the proof in the case of $Z=\ell^2_{\mathrm{per}}$, since the proof for the case of $Z=\ell^2_0$ is almost the same (note also that $\ell^p_0$ is a finite dimensional subspace of $\ell^p$). Due to  \eqref{DefLap_n}, we may apply \cite[Lemma A.1]{LLPhysD}, for the operators $\Delta_{(dn)}$ and $\Delta_{(dm)}$ respectively, to derive the inequality
\begin{eqnarray*}
	\|\Delta_dU \|_{\ell^p}^p
	&\leq&\sum_{n,m=0}^{N-1}\left(\vert\Delta_{(dn)}U_{n,m}\vert+\vert\Delta_{(dm)}U_{n,m}\vert\right)^p\\
	&\leq &c\sum_{n,m=0}^{N-1}\left(\vert\Delta_{(dn)}U_{n,m}\vert ^p+\vert\Delta_{(dm)}U_{n,m}\vert ^p\right)\\
	&\leq & c\left(c_1\|U \|_{\ell^p}^p+c_2\|U \|_{\ell^p}^p\right),
\end{eqnarray*}
 for some generic constants $c(p), c_i(p)>0$, $i=1,2$. 
\end{proof}

Inequality \eqref{genlp9} is proved in the last lemma.
\begin{lemma}
Let $Z={\ell^p_{\mathrm{per}}}, \ell^p_{0}$. The discrete biharmonic $\Delta^2:Z\rightarrow \ell^p$ satisfies the inequality $\|\Delta^2U\|_{\ell^p}\leq C\|U\|_Z$, for all $U\in Z$.
\end{lemma}
\begin{proof} In this case, we use \eqref{BDOp}, \cite[Lemma A.2]{LLPhysD} for the terms $\Delta^2_{(dn)}U_{n,m}$ and $\Delta^2_{(dm)}U_{n,m}$,  and Lemma \ref{LB1} for the terms $\Delta_{(dn)}(\Delta_{(dm)}U_{n,m})$ and $\Delta_{(dm)}(\Delta_{(dn)}U_{n,m})$, respectively. This way, we derive the inequality
\begin{eqnarray*}
	\|\Delta_d^2U \|_{\ell^p}^p
	&\leq& \sum_{n,m=0}^{N-1}\left(\vert\Delta_{(dm)}^2U_{n,m}\vert+\vert\Delta_{(dn)}(\Delta_{(dm)}U_{n,m})\vert+\vert\Delta_{(dm)}(\Delta_{(dn)}U_{n,m})\vert+\vert\Delta_{(dn)}^2U_{n,m}\vert\right)^p\\
	&\leq& c\sum_{n,m=0}^{N-1}\left(\left(\vert\Delta_{(dm)}^2U_{n,m}\vert+\vert\Delta_{(dn)}^2U_{n,m}\vert\right)^p+\left(\vert\Delta_{(dn)}(\Delta_{(dm)}U_{n,m})\vert+\vert\Delta_{(dm)}(\Delta_{(dn)}U_{n,m})\vert\right)\right)^p\\
	&\leq& c\sum_{n,m=0}^{N-1}\left(\vert\Delta_{(dm)}^2U_{n,m}\vert^p+\vert\Delta_{(dn)}^2U_{n,m}\vert^p+\vert\Delta_{(dn)}(\Delta_{(dm)}U_{n,m})\vert^p+\vert\Delta_{(dm)}(\Delta_{(dn)}U_{n,m})\vert^p\right)\\
	&\leq & c\left(c_1\|U \|_{\ell^p}^p+c_2\|U \|_{\ell^p}^p+c_3\|U \|_{\ell^p}^p+c_4\|U \|_{\ell^p}^p\right)=C\|U \|_{\ell^p}^p,
\end{eqnarray*}
again for some generic constants $c(p), c_i(p)>0$, $i=1,...,4$. 
\end{proof}

\end{document}